%% file: main.tex
\documentclass[12pt,a4paper]{article}

\usepackage[utf8]{inputenc}
\usepackage[british]{babel}
\usepackage{amsmath, amssymb, mathtools}

\usepackage{ltablex}
\keepXColumns

\usepackage[left=2.54cm, right=2.54cm, top=2.54cm, bottom=2.54cm]{geometry}
\usepackage{setspace}
\usepackage{fancyhdr}
\usepackage[authoryear,round]{natbib}
\setcitestyle{aysep={}}  

\usepackage{graphicx}
\usepackage{subcaption}
\usepackage{tcolorbox}
\usepackage[font=small,labelfont=bf]{caption}
\graphicspath{{./figures/}}

\usepackage{float}

\usepackage{booktabs}  
\usepackage{multirow}
\usepackage{array}
\usepackage{threeparttable}  
\usepackage{dcolumn}  

\newcolumntype{d}[1]{D{.}{.}{#1}}

\usepackage{xcolor}
\definecolor{darkblue}{rgb}{0,0,0.5}
\usepackage[colorlinks=true,
            linkcolor=black,
            citecolor=darkblue,
            urlcolor=darkblue,
            pdfborder={0 0 0}]{hyperref}

\usepackage{algorithm}
\usepackage{algpseudocode}

\usepackage{listings}
\usepackage{tikz}
\usetikzlibrary{shapes, arrows, positioning}

\tikzstyle{process} = [rectangle, rounded corners, 
                       minimum width=3cm, minimum height=1cm,
                       text centered, draw=black, fill=blue!10]
\tikzstyle{data} = [trapezium, trapezium left angle=70, trapezium right angle=110,
                    minimum width=2cm, minimum height=1cm,
                    text centered, draw=black, fill=green!10]
\tikzstyle{arrow} = [thick,->,>=stealth]

\usepackage[title,titletoc]{appendix}

\usepackage{chngcntr}
\AtBeginEnvironment{appendices}{%
  \counterwithin{figure}{section}
  \counterwithin{table}{section}
  \counterwithin{equation}{section}
}

\usepackage{eurosym}  
\usepackage{longtable}  
\usepackage{pdflscape}  

\usepackage{titling}
\newcommand\blfootnote[1]{%
  \begingroup
  \renewcommand\thefootnote{}\footnote{#1}%
  \addtocounter{footnote}{-1}%
  \endgroup
}

\usepackage{titlesec}

\titleformat{\section}
  {\normalfont\Large\bfseries}{\thesection}{1em}{}
  
\titleformat{\subsection}
  {\normalfont\large\bfseries}{\thesubsection}{1em}{}
  
\titleformat{\subsubsection}
  {\normalfont\normalsize\bfseries}{\thesubsubsection}{1em}{}

\newcommand{\Var}{\text{Var}}

\hypersetup{
    pdftitle={Interpreting the Interpreter: Can We Model post-ECB Conferences Volatility with LLM Agents?},
    pdfauthor={Umberto Collodel},
    pdfkeywords={monetary policy communication, large language models, central bank transparency},
    pdfsubject={Monetary Economics}
}

\begin{document}

\begin{titlepage}
\centering

{\LARGE\bfseries Interpreting the Interpreter: Can We Model Post-ECB Conference Volatility with LLM Agents?\par}

\vspace{1cm}

{\Large Umberto Collodel\par}

\vspace{0.2cm}

{\large Central Bank of Malta\\
Monetary Policy Department\par}

\vspace{0.2cm}

{\large \today\par}

\vfill

\begin{abstract}
\noindent
Central banks cannot observe market reactions to their communications before 
release. We propose a framework in which Large Language Models simulate 30 
heterogeneous traders interpreting European Central Bank press conference 
transcripts, yielding a measure of cross-sectional disagreement among synthetic 
agents. Across 293 Governing Council events from 1998 to 2026, this measure 
correlates at approximately 0.5 with realized Overnight Index Swap volatility, 
outperforming standard text-based alternatives in explaining market reactions. LLM-implied disagreement 
adds information beyond volatility clustering and remains robust in out-of-sample 
validation on genuinely unseen conferences from January 2025 onwards. We further 
show that providing historical examples of pre and post-conference volatility improves the calibration of model responses. The framework offers a practical tool for assessing, prior to release, how central bank communication is likely to be 
interpreted by financial markets.
\end{abstract}

\vspace{0.2cm}
\begin{flushleft}

\noindent\textbf{JEL Classification:} E52, E58, C63

\vspace{0.2cm}

\noindent\textbf{Keywords:}
monetary policy communication, large language models, 
monetary policy uncertainty, agent-based modeling
\end{flushleft}

\end{titlepage}

\blfootnote{The author would like to thank the instructors and participants of the 2025 Barcelona Graduate School of Economics Summer School on Natural Language Processing along with participants of the 2026 RCEA International Conference 
in Economics, Econometrics, and Finance. Special thanks go to Maximilian Freier for providing constructive feedback that greatly improved this work, as well as Manuel Bet\'in, Nicoletta Batini, Jonathan Benchimol, John Caruana, Francesco Toni, Eric Vaansteenberghe, Massimo Giovannini, and colleagues from the Central Bank of Malta. In particular, the work benefited from thorough discussions with Jacopo Zacch\`e. Any remaining errors are my own.  
This paper should not be reported as representing the views of the Central Bank of Malta. The views expressed are those of the authors and may not be shared by other research staff or policymakers in the Eurosystem without written permission by the authors. 

Correspondence to: Umberto Collodel \href{mailto:collodelu@centralbankmalta.org}({collodelu@centralbankmalta.org}).}

\newpage


\section{Introduction}\label{intro}

Central banks face a fundamental communication dilemma: they cannot test how  markets will interpret their statements before release. Upon publication, policy language triggers immediate and irreversible market reactions. Yet,  policymakers lack tools to anticipate these effects during the drafting process, when language remains modifiable. This constraint is costly. Ambiguous communication can amplify market volatility, hinder financial 
stability, and undermine policy transmission \citep{tillman2020,bauer2021market}. 
Traditional high-frequency identification methods operate ex-post, measuring 
what happened rather than guiding ex-ante language refinement (e.g. \citet{ALTAVILLA2019162}, \citet{acosta2025}).

This paper introduces an operational framework for simulating heterogeneous 
market reactions to monetary policy communication before release. We employ Large Language Models (LLMs) to simulate a cross-section of 30 heterogeneous synthetic traders, each endowed with distinct risk preferences, cognitive biases, and interpretive styles. These agents process European Central Bank (ECB) press conference transcripts and forecast Euro interest rate swap rates across three key maturities: 3-month, 2-year, and 10-year tenors. Cross-sectional forecast dispersion provides a model-based measure of market 
disagreement, which we validate against realized Overnight Index Swap (OIS) 
volatility across 293 ECB communications spanning June 1998 to March 2026.

Our analysis yields two main findings. First, simulated disagreement achieves Spearman correlations of approximately 0.5 with
realized market volatility for medium- and long-term maturities, peaking at
the 2-year tenor ($\rho$ = 0.53). LLMs thus appear to possess an intrinsic
capacity to decode central bank communication, even without prompt calibration,
historical conditioning, or model fine-tuning. Second, providing historical examples of post-conference volatility improves the calibration of model responses, reducing the bias and mean absolute error of the predictions, suggesting there is scope for further refinement of the framework.

We subject our results to four robustness checks targeting sampling stochasticity, 
look-ahead bias, construct validity, and volatility persistence. Across all 
specifications, the main results remain stable in magnitude, and statistical 
significance.  LLM-implied disagreement is further shown to be robust to prompt 
variation and specific model choice.

For central banks, the framework provides an operational tool to anticipate 
communication-induced volatility before release. Policymakers can 
systematically evaluate alternative phrasings and quantify expected market disagreement before publication. This 
transforms communication strategy from reactive refinement, based on costly 
market reactions, to proactive optimization during the drafting process. The 
methodology extends naturally to other major central banks, enabling 
comparative analysis of how communication styles and institutional frameworks 
affect interpretive disagreement across monetary policy contexts.

The remainder of the paper proceeds as follows. Section 2 reviews related 
literature on monetary policy communication, expectation formation, and LLM 
applications in economics. Section 3 describes our data sources, agent 
construction, and prompting strategies. Section 4 presents results on the 
correlation between simulated and market-based disagreement. Section 5 examines robustness of the results. Lastly, Section 7 concludes with implications for central bank communication design, caveats, and future research directions.

\section{Related Literature}

Our work contributes to three intersecting strands of research: monetary 
policy communication and high-frequency identification of market reactions, 
behavioral models of expectation formation, and the application of LLMs to economic analysis.

\paragraph{Monetary Policy Communication and High-Frequency Identification}

A substantial literature documents that central bank communication moves 
financial markets beyond the direct effects of policy rate changes. For the 
United States, \citet{kuttner2001} and \citet{gurkaynak2005actions} pioneered 
the identification of monetary policy shocks using high-frequency movements in 
federal funds rate and Eurodollar futures on Federal Open Market Committee 
(FOMC) dates. \citet{karadi2020} show that separating policy from information shocks in a structural VAR magnifies estimated monetary policy effects. \citet{bauer2023reassessment} 
further demonstrate that raw surprises are partially predictable from pre-announcement public 
data, recommending orthogonalization to restore instrument exogeneity. In the euro area, \cite{ALTAVILLA2019162} 
constructed a comprehensive database of monetary policy surprises from ECB 
announcements following the methodology of \cite{gurkaynak2005}. Similar 
strategies have been used to examine monetary policy in the United Kingdom 
\citep{CESABIANCHI2020103375}, China \citep{DAS2023102032}, and in a broad 
cross-country setting \citep{bolhuis2024}.

Subequent work by \citet{bauer2021market} distinguishes between the 
direction of policy surprises and the uncertainty surrounding the policy 
stance, showing that elevated uncertainty depresses equity prices, raises 
sovereign spreads, and increases exchange rate volatility. \citet{collodel2025market} 
document similar patterns for the Euro area, establishing that monetary policy 
uncertainty shocks have distinct transmission channels from conventional policy 
surprises. In addition, uncertainty might also diminish the transmission of first-moment surprises: for example, \citet{tillman2020} finds that hawkish surprises are less effective in periods of heightened monetary policy uncertainty, as investors shift toward 
longer maturities depressing the long-end of the curve.

While this literature convincingly demonstrates that communication matters, 
existing approaches remain fundamentally retrospective. High-frequency 
identification measures market reactions after communication occurs, offering 
no guidance for ex-ante language design. Our framework addresses this gap by 
enabling simulation of market disagreement before publication.

\paragraph{Expectation Formation and Behavioral Heterogeneity}

The rational expectations hypothesis, foundational to much of modern 
macro-finance, has been increasingly challenged by empirical evidence 
documenting systematic biases and bounded rationality among investors 
\citep{barberis2003survey,bordalo2018diagnostic}. Standard representative-agent 
models struggle to reconcile such behavioral deviations with observed phenomena 
like excess volatility and momentum. Agent-based models (ABMs) have emerged as 
a flexible alternative \citep{farmer2009economy,axtell2025agent}, allowing for 
heterogeneous, adaptive agents whose interactions generate emergent macroeconomic 
patterns. Within monetary policy research, ABMs have been used to model 
expectation formation and policy transmission in bottom-up settings 
\citep{delligatti2011macroeconomics}. Yet two key limitations remain. First, 
behavioral rules are typically imposed ex ante as stylized heuristics, rather 
than learned or emergent \citep{horton2023}. Recent work comparing heuristic 
and LLM-based agent architectures confirms that the latter substantially 
improve simulation realism in large-scale settings \citep{chopra2025limits}. 
Second, conventional ABMs cannot interpret qualitative information directly, 
despite the fact that monetary policy signals are largely communicated in 
natural language. This omission bypasses an essential channel of transmission. 
This distinction reflects a broader theoretical insight: while traditional 
prediction algorithms efficiently process observable data, they cannot access 
latent human judgment, a gap that generative AI begins is beginning to bridge by learning 
from large text corpora \citep{mullainathan2017machine,agrawal2018economics}. 
Our framework addresses both ABMs limitations by enabling agents to process and act 
upon policy-relevant textual information.

\paragraph{Large Language Models in Economic Analysis}

Recent work demonstrates that LLMs can simulate human decision-making in
economic contexts. \citet{horton2023} shows that LLM-based agents
(\textit{homo silicus}) replicate behavioral patterns in experimental market
settings. \citet{sinclair2025} extend this to institutional settings, simulating
FOMC deliberations through multi-agent systems and demonstrating that LLMs can
replicate collective decision-making processes. Most directly related to our
framework, \citet{hansen2025spf} construct a synthetic Survey of Professional
Forecasters panel by prompting LLMs with real-time macroeconomic data and
individual forecaster characteristics, showing that simulated forecasts replicate
key distributional features of human disagreement across horizons.

In central banking specifically, LLMs have been applied primarily to
\textit{classification} of communication content rather than prediction of market
reactions. \citet{imf2025cbcomms} use LLMs to classify 75,000 central bank
documents from 169 countries by topic and stance. \citet{bis2025cblm} develop
domain-adapted language models for hawkish-versus-dovish tone detection, while
\citet{centralbankroberta2023} build CentralBankRoBERTa to classify sentiment
toward macroeconomic agents. The Bundesbank's MILA system
\citep{bundesbank2024mila} employs role-based prompting to analyze ECB statements
sentence-by-sentence.

Our contribution departs from both strands. Unlike the LLM-agent literature,
which validates simulated behavior against experimental outcomes or survey
accuracy, we target  realized market outcomes.
Moreover, differently from the central banking classification literature, we are not interested in sentiment classification, but rather in how language generates heterogeneous interpretations. By combining these two angles, agent simulation and central bank communication, we produce a measure of
simulated disagreement that correlates with realized OIS volatility, enabling
ex-ante stress-testing of communication before market exposure. To our
knowledge, this is the first framework to use LLMs for prediction of market
disagreement following central bank communication.

\section{Methodology and Data}\label{methodology}

We construct a framework to simulate and predict market disagreement following 
ECB press conferences. The methodology comprises three components: data 
construction, agent design, and measure validation. We first assemble a comprehensive 
database of 293 ECB Governing Council press conferences spanning June 1998 to 
March 2026, pairing each transcript with realized Euro OIS 
volatility across the three maturities. We then develop LLM-based synthetic agents 
that process these transcripts and generate rate forecasts, with cross-sectional 
dispersion serving as our measure of simulated disagreement. Finally, we 
validate this measure by correlating it with actual market volatility.

ECB press conferences constitute the primary vehicle for monetary policy 
communication in the Euro area. Following the high-frequency identification 
literature \citep{ALTAVILLA2019162}, we define events 
as regularly scheduled Governing Council meetings. For each event, we collect 
full transcripts (Introductory Statement and Q\&A) and corresponding 
post-announcement OIS volatility for short-term (3-month), medium-term (2-year), 
and long-term (10-year) maturities.\footnote{Following standard practice, we 
use OIS rates as they reflect pure interest rate expectations without credit 
risk components present in government bond yields \citep{gurkaynak2005actions}.}

We proxy OIS volatility at day $t$ for maturity $m$ as the daily 
high-minus-low range on the day following the press conference:
\begin{equation}
\text{OIS Volatility}_{t,m} = \text{High}_{t+1,m} - \text{Low}_{t+1,m}
\end{equation}

The one-day window minimizes endogeneity concerns from confounding events e.g., macroeconomic data releases, geopolitical news, or other announcements, that would contaminate longer horizons. The high-low range provides a revealed-preference measure of belief dispersion: narrow ranges indicate consensus interpretations, while wide ranges reflect heterogeneous valuations driving active trading across the price spectrum.\footnote{We proxy standard deviation with the daily range due to intra-day tick data limitations.}\footnote{A widening of the range could, in principle, also reflect factors other than belief dispersion, such as liquidity conditions, balance-sheet constraints, or hedging demand. However, in complementary work \citep{collodel2025market}, we show that these alternative channels do not drive the gap.} We retrieve all quotes from LSEG Workspace.\footnote{Data for the 10-year tenor start in September 2011, compared to 2005 for the other two maturities.}

\begin{figure}[H]
    \centering
    \includegraphics[width=\textwidth]{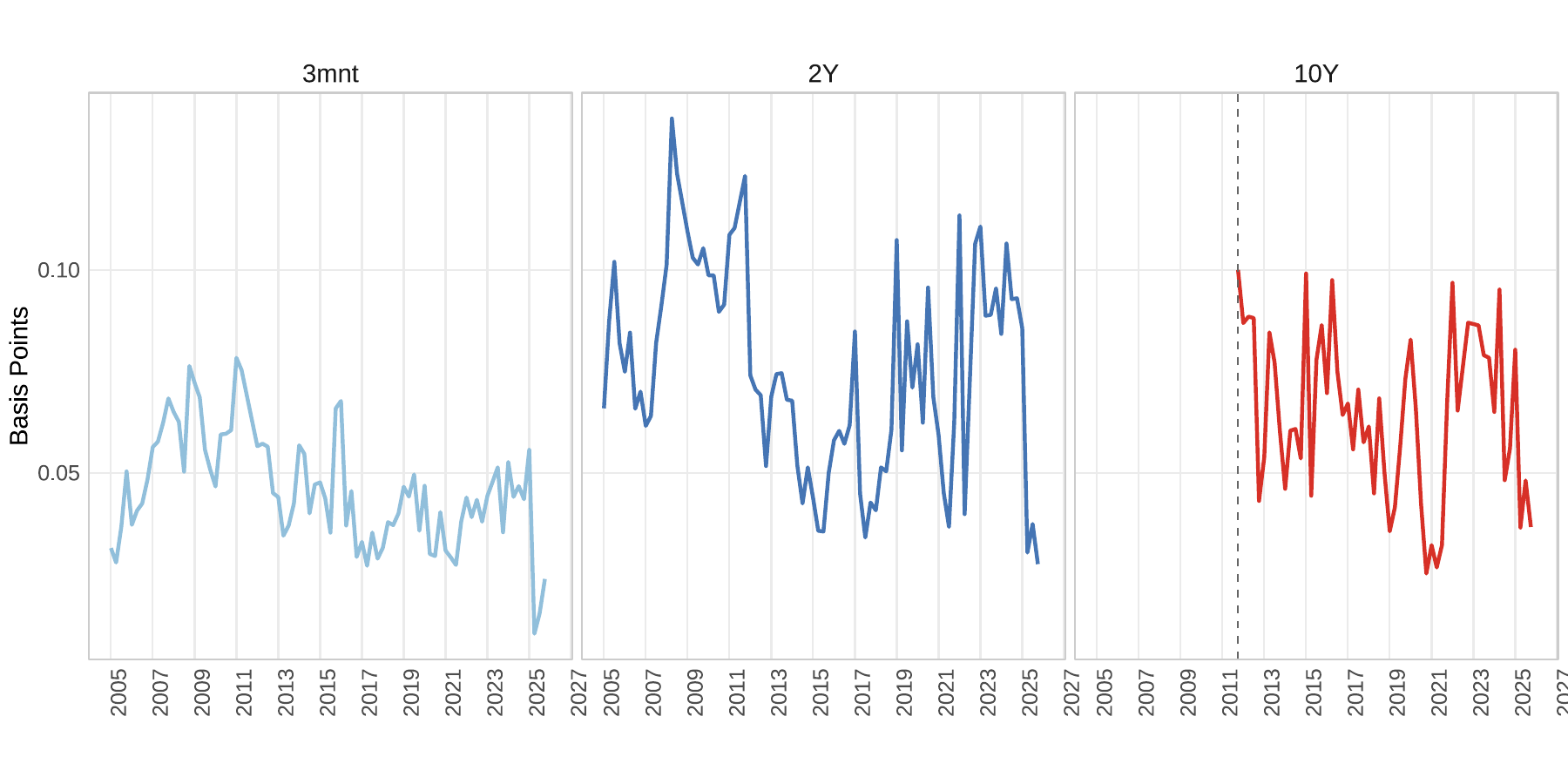}
    \caption{Post ECB Governing Council OIS volatility}
    \label{fig:actual_volatility}
    \scriptsize{\begin{flushleft}
    \textbf{Note}: Percentage points. Volatility is measured as the min-max range of OIS rates in the first day following ECB press conferences. Quarterly averages displayed. Data for the 10-year tenor available from 2011-Q3.
    \end{flushleft}}
\end{figure}

Figure \ref{fig:actual_volatility} illustrates market-based disagreement following ECB Governing Council meetings across all three maturities. The data reveal pronounced spikes during periods of heightened policy uncertainty, notably the 2008-2009 financial crisis, the European sovereign debt crisis, and the recent monetary policy tightening cycle initiated in 2022. Volatility exhibits a clear maturity structure, with consistently higher levels at longer tenors reflecting greater uncertainty about the medium- and long-term policy trajectory compared to near-term rate expectations.

\subsection{Agents}

We simulate 30 heterogeneous traders using Google's Gemini 2.5-Flash model.\footnote{We 
select 30 agents to balance computational efficiency with meaningful 
cross-sectional variation. Results are robust to using 20 or 50 agents and are available upon request.} Each agent 
interprets ECB press conference transcripts and forecasts Euro OIS rates across 
three maturities (3-month, 2-year, 10-year). To generate interpretive 
heterogeneity, agents differ along three dimensions:

\begin{enumerate}
    \item \textbf{Risk Aversion}: High, medium, or low tolerance for 
    uncertainty, affecting yield curve stability preferences.
    
    \item \textbf{Behavioral Biases}: Each agent exhibits 1--2 biases from 
    established behavioral finance (confirmation bias, overconfidence, anchoring, 
    herding, loss aversion, recency bias), reflecting systematic deviations 
    from rational expectations documented in real markets 
    \citep{barberis2003survey}.
    
    \item \textbf{Interpretive Style}: Agents adopt distinct analytical 
    approaches (fundamentalist, technical, sentiment-driven, quantitative, 
    narrative-focused, or policy-skeptic) that shape their processing of ECB conferences. These styles are drawn from heterogeneous agent models and behavioral finance (see \cite{HOMMES20061109}).
\end{enumerate}

The LLM dynamically assigns these characteristics to create 30 unique agents 
for each press conference, ensuring identical communication generates diverse 
interpretations. This design preserves behavioral diversity necessary to 
capture cross-sectional variation in market reactions while allowing natural 
adaptation over time. We set the model's temperature parameter to 1, 
balancing realistic heterogeneity with response stability.\footnote{Temperature controls 
output randomness: lower values (e.g., 0.2) produce deterministic responses, 
while higher values (e.g., 1.5+) generate implausible extremes.} However, since agent predictions are drawn stochastically at temperature 1, a single 
simulation run may not be representative of the underlying disagreement 
distribution. To address this, we repeat the simulation $R = 10$ times per 
press conference, computing the cross-sectional standard deviation of forecasts 
within each run and averaging across replications. The resulting 
$\widehat{\sigma}_t$ is therefore robust to the idiosyncratic noise of any 
individual draw, capturing the stable component of agent heterogeneity that 
systematic differences in trader characteristics generate rather than sampling 
variation.\footnote{See Section \ref{subsec:sampling} for details.}

It is important to note that these LLM-based agents represent stylized market 
participants who process textual information and form expectations. Real 
investors integrate broader factors, e.g. risk constraints, institutional mandates, collective behavior, that our framework abstracts from. This design isolates the communication-specific component of market disagreement, which is our target estimand, rather than attempting to match precisely realized price levels.

\subsection{Prompting Strategies}

We evaluate two prompting approaches in the main text:

\begin{description}
    \item[Baseline or Zero-Shot:] Agents receive only the ECB transcript. It tests whether LLMs possess intrinsic capabilities to decode monetary policy communication without explicit calibration.

    \item[Historical Anchoring or Few-Shot:] Augments the baseline by providing market volatility before and after the three previous press conferences. It tests whether historical context improves the accuracy of the predictions.
\end{description}

A third approach, reported in Appendix \ref{sec:judge}, further refines the prompting strategy:

\begin{description}
    \item[LLM-as-a-Judge:] A second LLM (``Judge'') iteratively reviews and 
    rewrites the prompt based on in-sample correlation with market 
    volatility, then validates optimized prompts out-of-sample.\footnote{We 
    use Gemini 2.5-Pro as the Judge and Gemini 2.5-Flash as the Analyst. 
    Train--test split is 75/25 in chronological order. See Algorithm 
    \ref{alg:llm_optimization} and Figure \ref{fig:judge-prompt} for details.}
\end{description}

For each event and prompting strategy, each agent generates forecasts of 
3-month, 2-year, and 10-year OIS rates, as well as the predicted direction 
and a confidence score.\footnote{We include the predicted direction to ensure internal consistency, as LLM outputs may otherwise exhibit inconsistencies.} 
Figure \ref{fig:llm-prompt} shows the naive prompt used in the first specification. 
We measure synthetic disagreement as the cross-sectional standard deviation of these forecasts across the 30 agents.

\subsection{Evaluation}

We evaluate the framework by correlating synthetic disagreement with realized 
market volatility. Our primary metric is the Spearman rank correlation 
coefficient, computed separately for each maturity. We use Spearman correlation 
because it captures monotonic relationships without linearity assumptions and 
exhibits robustness to outliers common in high volatility periods. We verify 
results using Pearson and Kendall correlations in Appendix 
\ref{subsec:correlation}.

For the LLM-as-a-Judge approach, we compare in-sample and out-of-sample 
performance to assess overfitting. We also examine maturity-specific patterns 
to identify which yield curve segments are most sensitive to communication-induced 
disagreement. Figure \ref{fig:methodology_pipeline} summarizes the complete 
framework.\footnote{For transparency and reproducibility, Appendix D provides 
implementation details including model specifications, text preprocessing, and 
computational cost estimates.}\footnote{For each press conference, we send a single API request. Although batching multiple conferences would significantly reduce computation time, since the request itself is the slowest part of the process, we adopt a conservative approach to avoid including future information as context, which could inadvertently influence the model’s output.} 
\begin{figure}[!htbp]
\centering
\caption{Stylized methodological pipeline for generating and validating synthetic disagreement from LLM-based agents}
\label{fig:methodology_pipeline}
\begin{tikzpicture}[node distance=0.6cm, auto, thick, scale=0.76, transform shape]
    \tikzset{
        data/.style = {rectangle, rounded corners, minimum width=3cm, minimum height=1cm, text centered, draw=black, fill=gray!20},
        process/.style = {rectangle, minimum width=3cm, minimum height=1cm, text centered, draw=black, fill=cyan!20},
        eval/.style = {rectangle, minimum width=4cm, minimum height=1cm, text centered, draw=black, fill=red!20}
    }
    \node (ecb) [data] {ECB Press Conference Transcript};
    
    \node (agents) [process, below=of ecb] {30 LLM-based Synthetic Agents};
    \node (prompt_few) [process, below=1.5cm of agents] {Few-Shot Prompting};
    \node (prompt_zero) [process, left=1cm of prompt_few] {Zero-Shot Prompting};
    \node (prompt_judge) [process, right=1cm of prompt_few] {LLM-as-a-Judge Prompting};
    
    \node (judge) [process, below=of prompt_judge] {Judge LLM (Gemini-Pro)};
    
    \node (forecasts) [data, below=1.5cm of prompt_few, align=center] {Rate Forecasts \\ (3m, 2y, 10y)};
    \node (synthetic) [process, below=of forecasts] {Compute Synthetic Disagreement};
    \node (market) [data, right=2.6cm of synthetic] {Market-Based Disagreement};
    
    \node (evaluation) [eval, below=2cm of synthetic] {Correlation Analysis \& Evaluation};
    \draw [->] (ecb) -- (agents);
    \draw [->] (agents) -- (prompt_zero);
    \draw [->] (agents) -- (prompt_few);
    \draw [->] (agents) -- (prompt_judge);
    \draw [->] (prompt_zero) -- (forecasts);
    \draw [->] (prompt_few) -- (forecasts);
    \draw [->] (prompt_judge) -- (forecasts);
    \draw [->] (forecasts) -- (synthetic);
    
    \draw [->] (synthetic) -- (evaluation);
    \draw [->] (market) -- (evaluation);
    
    \draw [->, dashed, red] (evaluation.east) -- ++(2.5cm,0) |- (judge.south);
    \draw [->, dashed, red] (judge) -- (prompt_judge);
\end{tikzpicture}
\end{figure}


\section{Results} \label{sec:results}

This section evaluates the performance of our two main prompting strategies in replicating market disagreement patterns following ECB press conferences. We examine the baseline (Section \ref{sec:zero_shot}) and the few-shot approach with historical context (Section \ref{subsec:historical_anchoring}). Results for the iterative LLM-as-a-Judge framework are in Appendix \ref{sec:judge}).

\subsection{Baseline}\label{sec:zero_shot}

\begin{figure}[!htbp]
  \centering
  \includegraphics[width=0.65\textwidth]{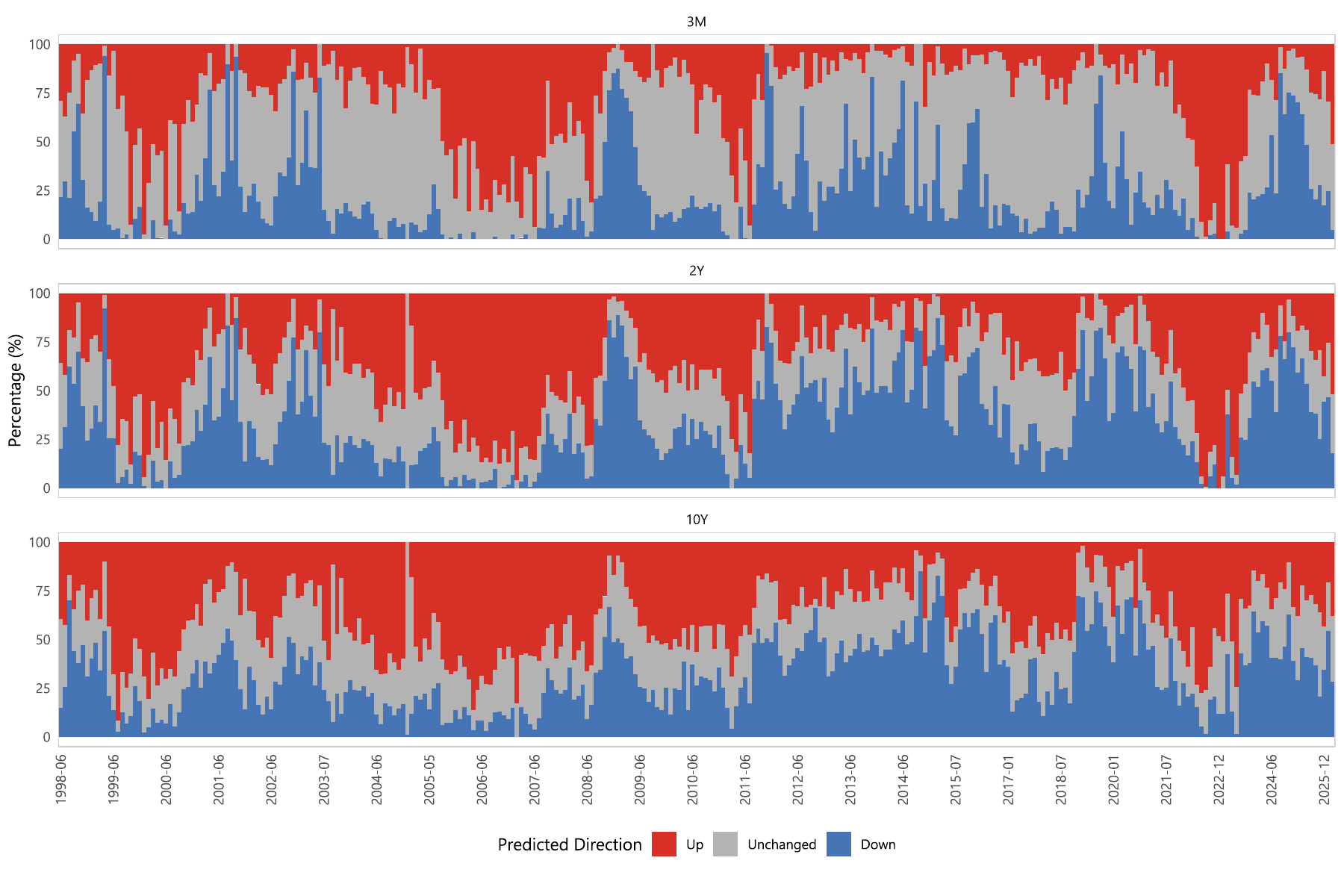}
  \caption{Percentage of forecast directions from the zero-shot LLM simulation}
  \label{fig:direction_zero}
  \vspace{1ex}
  \scriptsize\begin{flushleft}
  \textbf{Note:} Each data point reflects the output of an LLM model prompted in a zero-shot setting using Gemini 2.5-Flash, with temperature 1. The simulation uses a sample of 293 ECB press conference (June 1998--March 2026) transcripts and the baseline prompt (Figure \ref{fig:llm-prompt}). For each press conference, the model output is generated across 10 independent sampling runs.
  \end{flushleft}
\end{figure}

\begin{figure}[!htbp]
  \centering
\caption{Standard deviation of forecasts from the baseline LLM simulation}
  \label{fig:ts_zero}
  \includegraphics[width=0.8\textwidth]{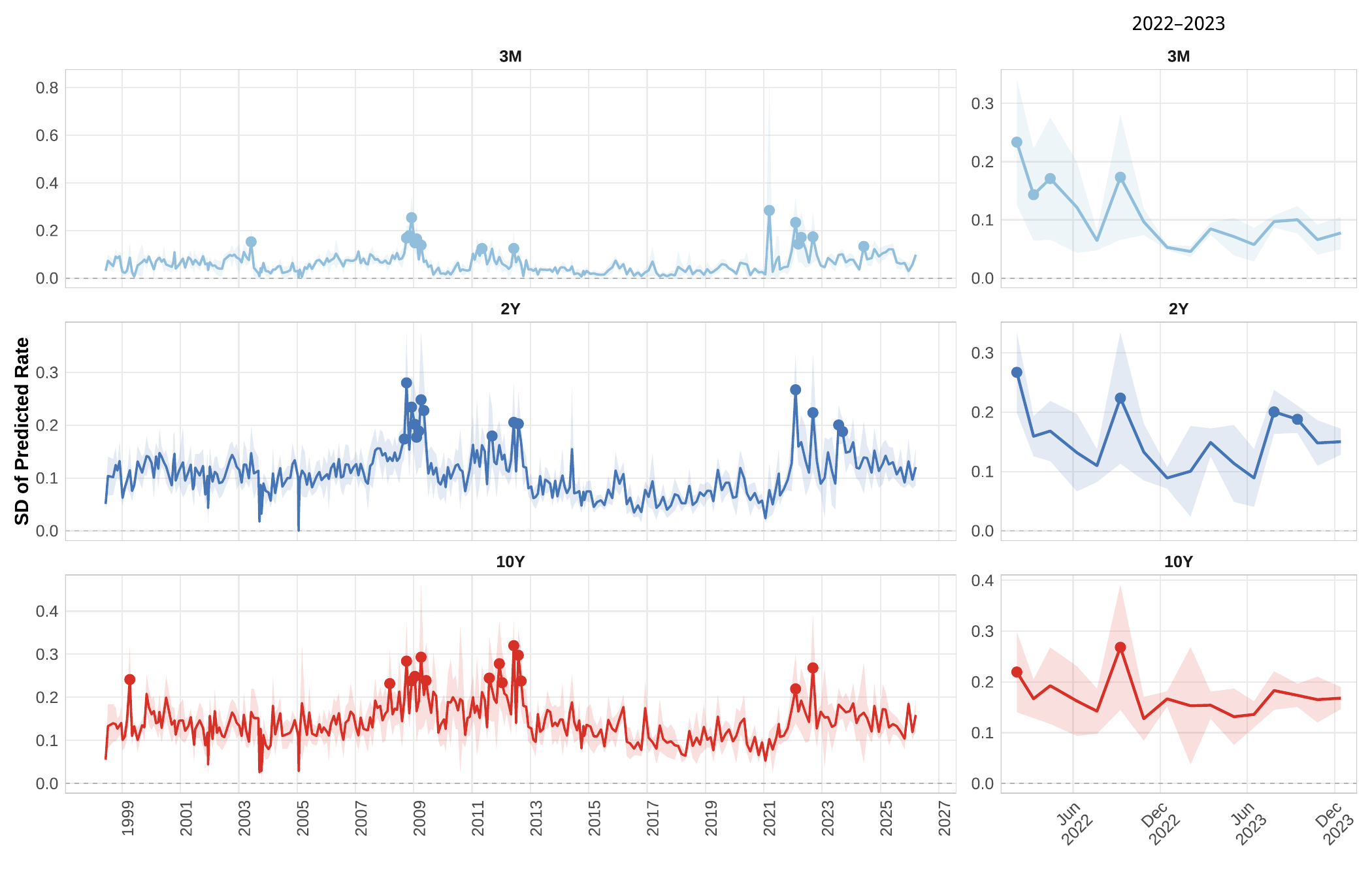}
  \scriptsize\begin{flushleft}
  \textbf{Note:} Each data point reflects the output of an LLM model prompted in a zero-shot setting using Gemini 2.5-Flash, with temperature 1. Dispersion is measured as the average cross-sectional standard deviation of rate predictions over 10 iterations. The simulation uses a sample of 293 ECB press conferences (June 1998--March 2026) and the baseline prompt (Figure \ref{fig:llm-prompt}). Highlighted points represent observations exceeding the 95th percentile of dispersion for each tenor. Shaded areas are 95\% confidence intervals.
  \end{flushleft}
\end{figure}

The baseline specification provides the most direct test of whether LLM agents endowed only with a pool of behavioral traits to pick from and no historical training context can reproduce the ebb and flow of overall market sentiment. We run the model at temperature 1 and average outputs across 10 independent sampling runs per press conference to stabilize the disagreement signal without sacrificing the interpretive nuance of the underlying text. Before turning to formal correlation measures, Figures \ref{fig:direction_zero} and \ref{fig:ts_zero} offer a visual narrative of the simulated output.

Figure \ref{fig:direction_zero} plots the share of agents predicting an increase, decrease, or no change in rates after each ECB press conference. Even in this naive setup, the agents’ directional calls broadly mirror well-known phases of ECB policy: a tightening phase before the Great Recession, a pronounced easing wave 
thereafter, a prolonged "on hold" stance at the short end from the zero 
lower bound (ZLB) through the pandemic (while long maturities drifted 
steadily lower), followed by the rapid tightening and partial reversal of 
the recent cycle.

Figure \ref{fig:ts_zero} shows our masure of interest: the cross-sectional standard deviation of forecasts.

Dispersion contracts systematically during periods of unambiguous policy stance, such as the post-crisis easing phase and much of the ZLB era, when forward guidance provided clear anchoring. In this period, the 3-month dispersion falls below 5 basis points, with longer tenors, instead, clustering around 10-15 basis points. Conversely, it spikes at policy turning points and during crisis episodes, peaking in the 2008-2009 financial crisis (with the 2 and 10-year dispersion exceeding 40 basis points), the 2011-2012 sovereign debt crisis, and most recently during the 2022-2023 monetary policy tightening cycle, where disagreement reaches unprecedented levels (roughly 60 basis points for the 2 and 10-year maturity). In the latter instance, disagreement arises less from directional confusion and more from uncertainty over policy speed, magnitude, and terminal endpoints. This last result is consistent with the ECB 2025 Monetary Policy Strategy Review findings \citep{ECB_OP372_2024}. The report shows that, once forward guidance was abandoned, the sensitivity of OIS rates to macroeconomic news increased significantly, hence amplifying excess volatility. Moreover, the early phase of the 2022 hiking cycle was marked by historically large monetary policy shocks, including the unexpected 50 basis point hike in July 2022 and the largest hawkish surprise since the global financial crisis in December 2022, reflecting the challenge for the ECB of aligning communication with policy moves. 
By contrast, following the introduction of the three-element reaction function framework in March 2023, the subsequent phase of the tightening cycle exhibits markedly subdued volatility to policy announcements.\footnote{These patterns mirror survey-based and market-implied measures of disagreement, where even directionally well-telegraphed policy paths can generate substantial heterogeneity in rate forecasts. See Gilchrist et al. (2014) and Bayer et al. (2022) for a thorough discussion on forecast disagreement and monetary policy uncertainty.}\footnote{At the end of 1998, volatility exhibits another peak for the 3-month tenor. In this period, the ECB avoided signaling its initial policy rate for the monetary union, deferring decisions to December. While acknowledging global risks from the Asian crisis, Russian default, and LTCM collapse, President Duisenberg reaffirmed the commitment to price stability, contrasting with the Fed’s three rate cuts between September and November. This ambiguity left markets uncertain and drove volatility in short-term instruments, mainly 1- to 3-month FRAs and DEM-based futures, since euro derivatives like Euribor futures and OIS did not yet exist \citep{ecb1998annual}.} The cross-tenor hierarchy remains remarkably consistent across all episodes, with 10-year uncertainty slightly larger than 2-year uncertainty, and both larger than the 3-month dispersion by 2-3 times, reflecting the increasing complexity of information processing and projection uncertainty at longer time horizons.

\begin{figure}[H]
    \centering
        \caption{Mean Spearman correlation with market-based measures by tenor}
    \label{fig:mean_spearman_correlation}
    \includegraphics[width=0.66\textwidth]{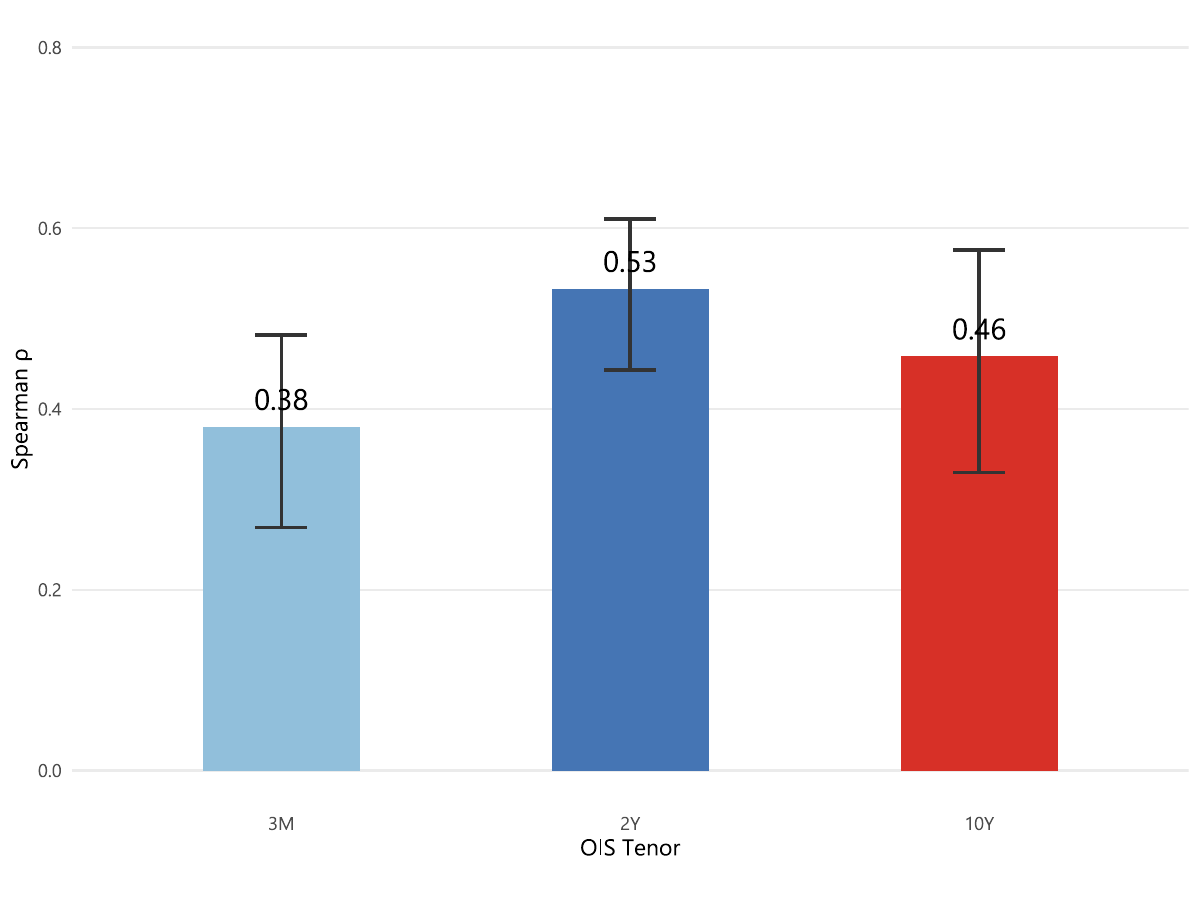}
    \vspace{1ex}
    \scriptsize\begin{flushleft}
    \textbf{Note:} Correlation between the output of an LLM model prompted in a zero-shot setting using Gemini 2.5-Flash, with temperature 1, and the OIS min-max range 1 day after ECB conferences. The simulation uses a sample of 293 ECB press conferences (June 1998--March 2026) and the baseline prompt (Figure \ref{fig:llm-prompt}). 90\% confidence intervals computed via a non-parametric bootstrap with 5{,}000 replications. For each press conference, the model output is generated across 10 independent sampling runs and then averaged to ensure robustness.
    \end{flushleft}
\end{figure}

Figure \ref{fig:mean_spearman_correlation} presents correlations with market measures. The 2-year tenor achieves the highest correlation (0.53), followed by the 10-year (0.46), and the 3-month (0.38). This hierarchy likely reflects the differential nature of information processed by the LLM across the yield curve. The 2-year tenor primarily captures short- to medium-term policy expectations that are directly shaped by ECB communication during press conferences. In contrast, the 10-year yield incorporates broader factors extending beyond central bank communication, e.g., long-term inflation expectations, structural growth prospects, and term premium fluctuations, making it less tightly linked to the specific content of ECB press conferences. The 2-year segment's narrower focus on monetary policy expectations, therefore, produces the strongest correlation with our LLM-based disagreement measure. At the short end, the 3-month OIS rate exhibits inherently low variance, particularly during prolonged policy stasis such as the ZLB era. Consequently, even when the LLM accurately interprets ECB communication tone, the constrained variation in 3-month rates limits the achievable correlation. Crucially, these results demonstrate that even a simple zero-shot LLM approach, without historical training, prompt-optimization, or fine-tuning, can capture economically meaningful patterns of market disagreement following central bank communication.

To gauge whether this performance can be replicated by off-the-shelf text metrics, we compare our LLM-based disagreement measure against five straightforward textual benchmarks. Complexity measures include Flesch--Kincaid readability scores and document length (word count), capturing syntactic structure and information volume. Framing and tone measures comprise hedge word density and Loughran--McDonald uncertainty terms, which proxy for linguistic ambiguity and explicit uncertainty language. Finally, stance measures capture the net balance of hawkish versus dovish terminology, reflecting directional policy tone.\footnote{Flesch--Kincaid Grade Level estimates the U.S. school grade required to comprehend a text based on average sentence length and syllables per word; higher values indicate more complex, technical language. Word count measures total document length, proxying for informational load. Hedge word density captures the frequency of cautious or non-committal expressions (e.g., may,'' could,'' possibly,'' somewhat'') that signal reduced commitment or linguistic uncertainty. Loughran--McDonald uncertainty comprises terms from the finance-specific dictionary developed by \cite{loughran2011}, designed to identify explicit uncertainty expressions in economic and financial contexts (e.g., uncertain,'' risk,'' volatility''). The net hawkish--dovish score measures the balance between hawkish terms (e.g., inflation,'' tightening,'' overheating'') and dovish terms (e.g., slowdown,'' accommodation,'' ``support''), following standard monetary policy dictionaries.}  In related work, \cite{mumtaz2023keep} show that Bank of England announcements with lower readability score are associated with higher gilt yield volatility post-announcement.

Table~\ref{tab:complexity_correlations} reports Spearman correlations
between these benchmarks and post-announcement disagreement. All five
exhibit weaker correlations than our LLM ensemble across all tenors,
and the gap widens at longer maturities: at the 10-year tenor, our
measure achieves 0.46, while the strongest text-based benchmark
(Hedging Words) reaches roughly 0.19.

\input{tables/robustness/complexity_correlations}

The pattern is informative about what these benchmarks measure.
Each is conceptually adjacent to communication-induced
disagreement but distinct from it: for example, a transcript can be syntactically
clear yet interpretively ambiguous. These textual proxies surface features of
text; our measure proxies the act of interpretation itself. This distinction is consistent with the maturity gradient. Surface
metrics retain modest correlation with 3-month volatility, where
near-term policy decisions are anchored to the press conference's
explicit statements. They lose traction at longer maturities, where
disagreement reflects how markets reconcile forward guidance with
the broader policy reaction function, content that is invisible to
metrics scoring sentence length or hedge counts.

Taken together, the zero-shot results establish a measurable, content-sensitive
link between ECB communication and post-announcement market disagreement. 

\subsection{Historical Anchoring}
\label{subsec:historical_anchoring}

We next turn to the role of additional information in the prompt. \cite{hansen2025spf} show that incorporating lagged median forecasts from SPF survey participants improves an LLM’s ability to replicate forecasts in subsequent survey rounds. This finding is consistent with the few‑shot learning literature, which shows that providing relevant examples in the prompt can improve predictive accuracy without incurring in the computational cost associated with updating model parameters \citep{brown2020fewshot}.

\begin{figure}[H]
\centering
\caption{Mean Spearman correlation with market-based measures by tenor}
\label{fig:few_shot}
\includegraphics[width=0.75\textwidth]{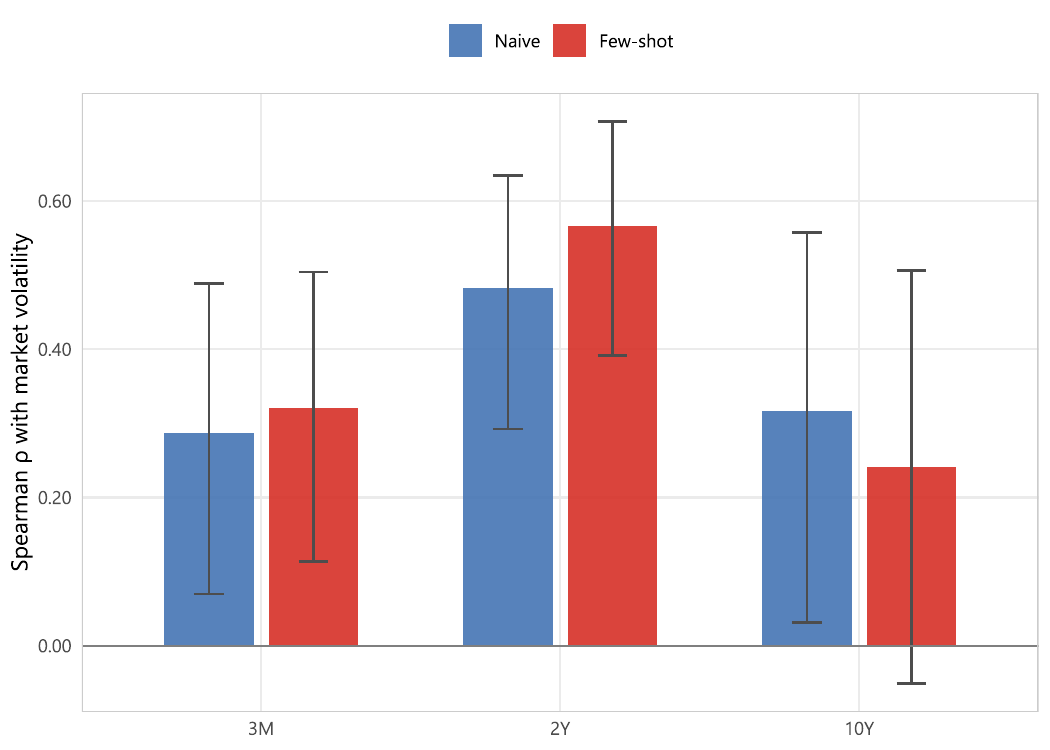}
\vspace{1ex}
\scriptsize
\begin{flushleft}
\textbf{Note}: Correlation between the output of an LLM model prompted in a few-shot setting using Gemini 2.5-Flash, with temperature 1, and the OIS min-max range 1 day after ECB conferences. The simulation uses a subset of 90 ECB press conferences, 30 drawn from each tercile of realized post‑conference OIS volatility at the 2‑year maturity, from the original sample of 293 ECB press conferences (June 1998--March 2026) and the prompt with historical anchoring (Figure \ref{fig:llm-prompt-historical}). 90\% confidence intervals computed via a non-parametric bootstrap with 5{,}000 replications. For each press conference, the model output is generated across 10 independent sampling runs and then averaged to ensure robustness.
\end{flushleft}
\end{figure}

To assess this mechanism in our setting, we augment the baseline prompt with the realized before‑ and after‑conference OIS min–max range from the three most recent meetings, computed separately for each maturity. The intuition is straightforward: a trader operating in financial markets observes not only the content of the ECB’s communication, but also the volatility environment surrounding recent meetings. The few‑shot prompt seeks to replicate this informational context. As in the previous section, we then compute Spearman correlations with market‑based measures.\footnote{We run the simulation on a subset of 90 ECB press conferences, 30 drawn from each tercile of realized post‑conference OIS volatility at the 2‑year maturity, in order to minimize computational cost without sacrificing statistical power.}\footnote{As in the previous section, we average the standard deviation for each conference across 10 runs to erase sampling noise and allow comparison.} Figure~\ref{fig:few_shot} reports the results.

In terms of ranking i.e., which conferences generate more or less disagreement, the two prompting strategies are statistically indistinguishable. The Spearman correlation between the benchmrk and few‑shot disagreement series exceeds 0.7 at all maturities, and differences in their respective correlations with realized market volatility are never statistically significant. Both prompts identify the same press conferences as high‑ or low‑uncertainty events, suggesting that this signal is intrinsic to the model’s interpretation of the transcript rather than dependent on historical anchoring.\footnote{Confidence intervals for the 10-year maturity are particularly large due the sample size.}

Where the two approaches diverge sharply, however, is in the level of predicted disagreement. Table~\ref{tab:mae} reports the resulting biases and mean absolute errors for the baseline and the specification with anchoring.

\input{tables/few-shot/fewshot_vs_naive_calibration}

The baseline systematically over‑predicts realized volatility across all maturities: the estimated bias is positive and statistically significant throughout, with the distortion particularly pronounced at the 10‑year tenor. The few‑shot prompt largely corrects this bias: estimates become small and, at the 10‑year maturity, statistically indistinguishable from zero. Correspondingly, the mean absolute error declines substantially, by roughly 19 percent at the 2‑year and 41 percent at the 10‑year maturity, when historical context is included.

Taken together, these results suggest a natural division of labor between the two measures. For applications that require only the direction of uncertainty, such as identifying episodes of heightened disagreement or ranking press conferences by their informational content, the baseline prompt performs well and avoids the additional data requirements of the few‑shot approach. For applications that require accurate level estimates of market volatility, such as using the synthetic measure as a direct input into a risk model or a communication assessment tool, the few‑shot prompt offers markedly better calibration.

\section{Robustness}
\label{sec:robustness}

We subject the baseline results to four robustness checks targeting the principal threats to validity in our setting. First, we assess whether the LLM disagreement measure is sensitive to sampling stochasticity: since agent predictions are drawn at temperature 1, repeated simulation runs may yield different cross-sectional dispersions for the same transcript. Second, we address look-ahead bias by conducting a strict out-of-sample test using exclusively press conferences that post-date Gemini 2.5-Flash's knowledge cut-off, ensuring the model processes genuinely unseen transcripts. Third, we examine a construct validity threat specific to the agent-based design: because panel generation and rate forecasting occur within the same prompt in the baseline, the LLM may implicitly condition the characteristics of the 30 synthetic traders on the transcript it has already processed. Thus, the model would manufacture dispersion top-down through panel construction rather than allowing it to emerge organically from heterogeneous processing of a common signal. We address this through a two-stage experiment that separates panel generation, conducted using only the date of the conference, from forecast elicitation. Fourth, we control for volatility persistence by augmenting the baseline specification with pre-conference OIS volatility, testing whether the LLM measure retains predictive power beyond autoregressive dynamics. 

\subsection{Stochasticity and Replicability}
\label{subsec:sampling}

 Since the model generates responses stochastically, running the same
  press conference through the simulation twice yields two different sets
   of trader forecasts and, consequently, two different disagreement     
  estimates. Embedding stochasticity in the model is a deliberate choice as it allows the       
  synthetic panel to explore the full range of plausible interpretations 
  of a given communication rather than collapsing to a single
  deterministic output, but it also implies that a single simulation draw is 
  a noisy estimate of the underlying conference-level disagreement.  If   
  that noise were large relative to the signal of interest, the
  correlations with realised market volatility reported in Section~\ref{sec:zero_shot}     
  could be sensitive to the particular draw used. In addition, stochasticity hinders external replicability:
independent researchers re-running the simulation on the same transcripts will
obtain numerically different values.

To address both concerns, we repeat each simulation $R = 10$ times and average the standard deviation per conference across draws. In principle, the
resulting ensemble disagreement measure should be more stable than any individual
draw: idiosyncratic sampling variation cancels across runs, while the
conference-specific signal accumulates. Moreover, in this way approximate
replicability is preserved: independent researchers running a sufficient
number of draws will obtain ensemble averages that converge to the same
underlying conference-level signal, even if individual draws differ.

We assess how much averaging is sufficient using a reliability
  coefficient adapted from psychometrics. For conference $i = 1, \ldots, N$ and run $r = 1, \ldots, R$, let $d_{i,r}$ be the disagreement estimate, and $\bar{d}_i = \frac{1}{R}\sum_{r=1}^{R} d_{i,r}$ the conference mean across runs.

$\sigma^2_{\text{within}}$ is computed for each conference as the sample variance of its $R$ draws around its own mean, then averaged across all conferences:

\begin{equation}
\hat{\sigma}^2_{\text{within}} = \frac{1}{N} \sum_{i=1}^{N} \frac{1}{R-1} \sum_{r=1}^{R} (d_{i,r} - \bar{d}_i)^2
\end{equation}

$\sigma^2_{\text{between}}$ is the sample variance of the conference means $\bar{d}_i$ across the $N$ conferences:

\begin{equation}
\hat{\sigma}^2_{\text{between}} = \frac{1}{N-1}\sum_{i=1}^{N}(\bar{d}_i - \bar{\bar{d}})^2
\end{equation}

where $\bar{\bar{d}} = \frac{1}{N}\sum_{i=1}^{N}\bar{d}_i$ is the grand mean.
  
\begin{figure}[H]
  \centering
    \caption{Reliability per Runs Averaged}
  \label{fig:gr_stabilization}
  \includegraphics[width=0.8\textwidth]{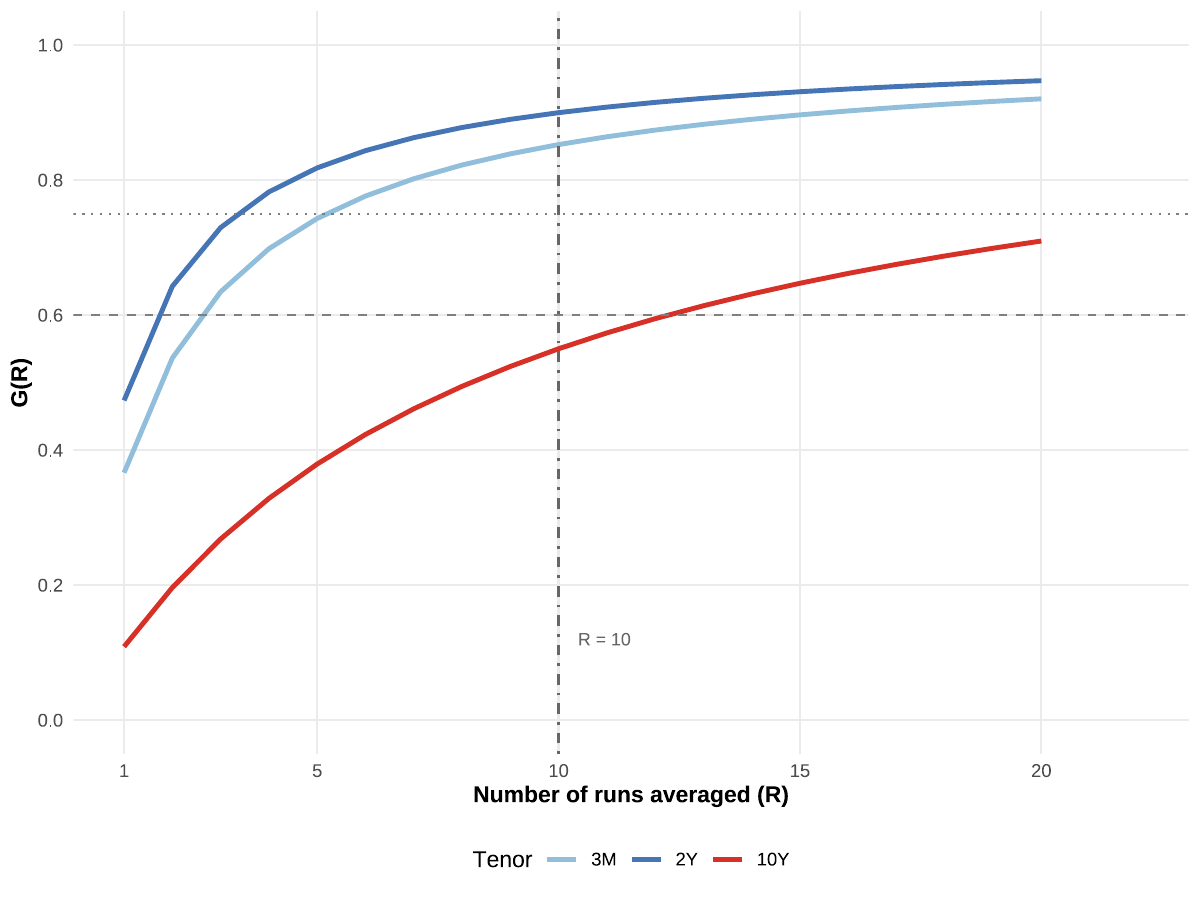}
  \vspace{1ex}
  \scriptsize\begin{flushleft}
  \textbf{Note:} $G(R)$ is the analytical reliability curve with $\hat{\sigma}_{\text{between}}$ and      
  $\hat{\sigma}_{\text{within}}$ estimated from the full $R=10$ ensemble; the entire curve evaluates the
  formula at hypothetical run counts, with the vertical marker indicating the observed ensemble size. Values for $R< 10$ and $R >10$ are back-projected and extrapolated, respectively. Dashed and dotted horizontal lines indicate 0.60 (good) and 0.75 (excellent) reliability thresholds                 
  \citep{cicchetti1994}.
  \end{flushleft}
\end{figure}

The reliability of an $R$-run ensemble is then:     

  $$G(R) = \frac{\sigma^2_{\text{between}}}{\sigma^2_{\text{between}} +  
  \sigma^2_{\text{within}} / R},$$

  which ranges from 0 (dominance of sampling noise) to 1 (perfect signal recovery).\footnote{The formula is the Spearman--Brown prophecy formula 
\citep{spearman1910,brown1910}, originally derived in psychometrics to 
predict the reliability gain from lengthening a test by a factor of $R$.} \citet{cicchetti1994} provides conventional benchmarks for interpreting 
$G$: values below 0.40 indicate poor reliability, 0.40--0.59 fair, 0.60--0.74 good, 
and 0.75 or above excellent.      
  Figure~\ref{fig:gr_stabilization} plots  the analytical $G(R)$ for $R = 1, \ldots, 20$ across the    
  three tenors.\footnote{Although in principle we could construct the empirical function, subsetting our sample for each value of R,  the estimates at values lower than 10 would be very unstable.} 
  
  A single draw achieves reliability in the range of       
  0.1--0.5, whereas the $R = 10$ ensemble reaches 0.75, on average, comfortably above 
  the conventional 0.60 threshold and comparable in magnitude to measurement in high-quality administrative surveys (e.g. \cite{Pischke1995}, \cite{SchmillenUmkehrerWachter2024}). Moreover, gains per additional run diminish rapidly, suggesting that $R = 10$ sits near the point of practical convergence. 
  
\begin{figure}[!htbp]
  \centering
    \caption{Stabilization of Disagreement Estimates over Different Runs}
  \label{fig:sd_stabilization}
  \includegraphics[width=0.95\textwidth]{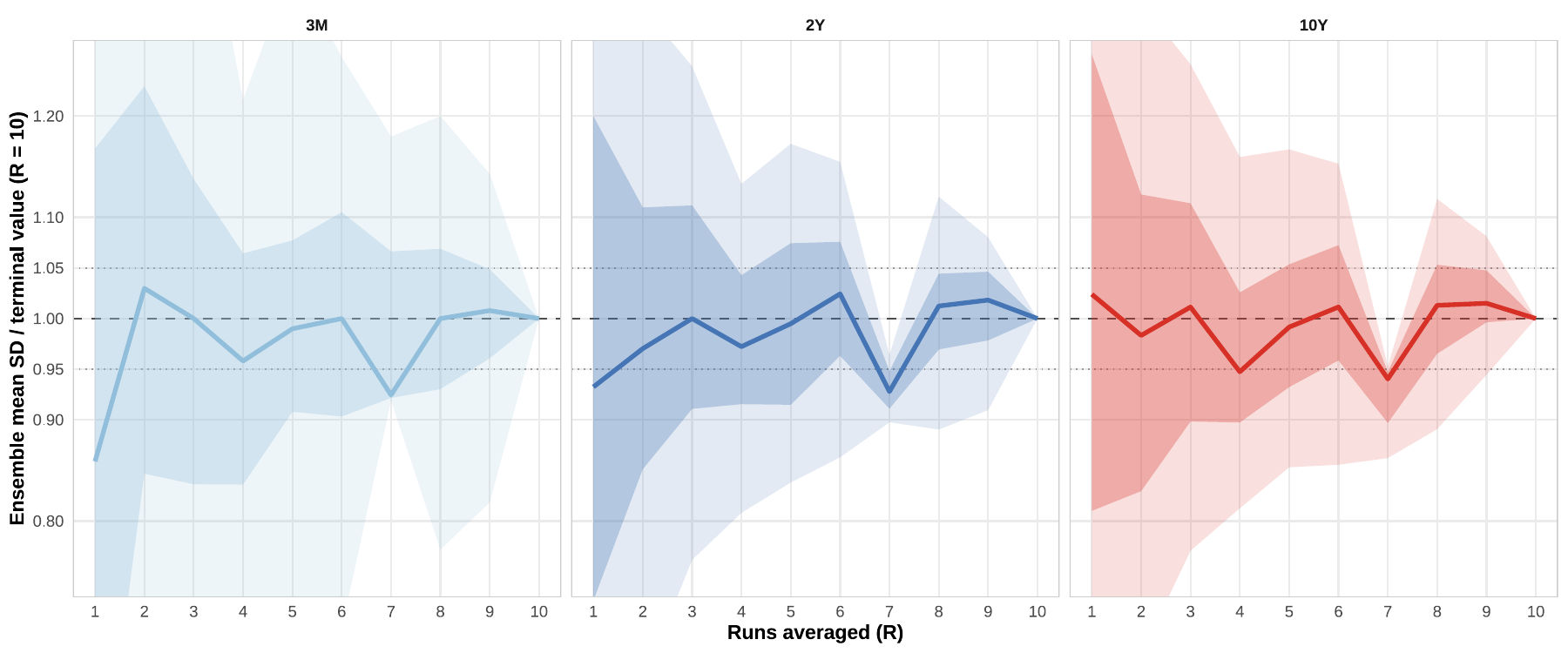}
  \vspace{1ex}
  \scriptsize\begin{flushleft}
  \textbf{Note:} For each tenor, the plot reports the distribution across press conferences of the ensemble mean cross-sectional standard deviation at run count $R$, normalized by its value at $R=10$ ($R=10$ equals 1). Dark band shows the interquartile range; light band shows the 5th--95th percentile range; the line shows the median. Dotted lines indicate $\pm 5\%$ around the $R=10$ value.
  \end{flushleft}
\end{figure}

Figure~\ref{fig:sd_stabilization} corroborates this evidence directly: the mean across runs of individual conferences' standard deviation, normalised by its value at $R = 10$, stabilises to within $\pm 5\%$ of the terminal estimate by the seventh run for the bulk of the distribution.

The two diagnostics establish that averaging at $R = 10$ provides a stable and approximately 
replicable measure of conference-level disagreement. Approximate replicability 
holds because each simulation draw is an independent realisation conditional on 
the transcript: independent researchers running a sufficient number of draws will 
obtain ensemble averages that converge to the same conference-level signal, even 
if individual draws differ. We conclude that the correlations with realised OIS volatility reported in Section~\ref{sec:zero_shot} are therefore robust to the sampling parameter.

\subsection{Look-Ahead Bias and Out-of-Sample Validation}

A key concern for any LLM-based empirical approach is whether the model genuinely processes textual content or merely exploits patterns memorized during training (e.g. \citet{lopezlira2025memorization}). Since the version of the Gemini 2.5-Flash model we use is trained on data up to January 1, 2025, conferences in our sample from 1998 to 2024 might be affected. To address this methodological concern, we implement several mitigation strategies. First, we explicitly instruct the model to condition their responses on information available only up to the specified conference date, with prompts that emphasize adherence to historical information boundaries \citep{hansen2025spf}. This approach attempts to replicate the informational constraints faced by actual market participants at each point in time.\footnote{From a technical standpoint, it lowers the probabilities of next-word tokens being the ones that include future information.} Second, from a technical standpoint, we process each press conference through individual API requests rather than batching multiple conferences together. While batching would significantly reduce computational time, it would risk contaminating earlier forecasts with information from subsequent conferences within the same batch. Our sequential, single-conference approach ensures that the model's context window contains only the target conference and its associated prompt, preventing any inadvertent temporal spillover between events.

Nevertheless, while these safeguards reduce the risk of temporal leakage within the training sample, they cannot fully address the possibility that analyses of ECB communication were included in the model's training corpus. To provide definitive evidence that our results reflect genuine textual processing rather than memorization, we conduct a strict temporal out-of-sample test using only conferences that post-date the model's training period. Specifically, we re-run the LLM simulation exclusively on ECB press conferences held between January 2025 and March 2026, a period entirely outside the model's knowledge cutoff. This design ensures the model processes genuinely unseen transcripts, providing a clean test of generalization. While the out-of-sample period is necessarily limited, this represents the strongest possible validation of our approach.

\begin{figure}[!htbp]
    \centering
    \includegraphics[width=0.75\textwidth]{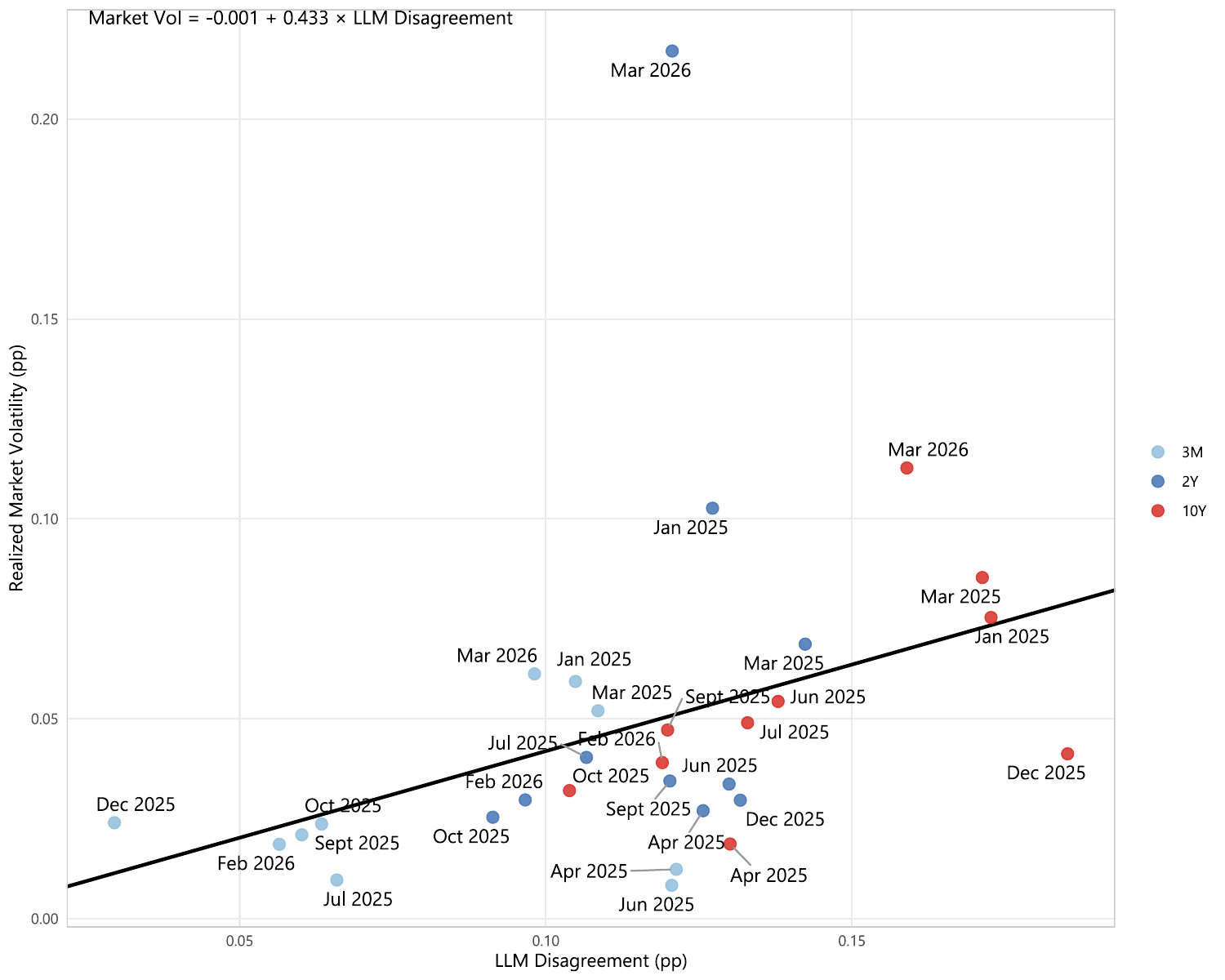}
    \caption{Out-of-Sample Performance --- LLM Disagreement vs Market Volatility}
    \label{fig:oos_validation}
    \scriptsize\begin{flushleft}
        \textbf{Note:} Each point represents an ECB press conference post January 1, 2025 (after Gemini 2.5-Flash knowledge cut-off). Synthetic disagreement is measured as the cross-sectional standard deviation of rate predictions across 30 heterogeneous agents generated using Gemini 2.5-Flash with the zero-shot prompt (Figure \ref{fig:llm-prompt}) and temperature 1. Market volatility is the min-max range of OIS rates 1 day after ECB press conferences. For each press conference, the model output is generated across 10 independent sampling runs and then averaged to ensure robustness.
\end{flushleft}
\end{figure}

Figure \ref{fig:oos_validation} presents the results, pooling observations across all three tenors. 

The model's performance on this genuinely unseen data is, as expected, inferior to the in-sample exercise, but, nevertheless, synthetic disagreement exhibits high correlation with market values: 0.43 across 30 tenor-date observations.

One observation warrants closer examination: the March 2026 conference registers markedly elevated realized volatility relative to the synthetic disagreement generated by the model, especially so for the 2 and 10 years maturities. The outsized market reaction that day was driven primarily by the sudden escalation of the Middle East conflict. Our LLM agents, reading only the communication, correctly assessed it as high ambiguity, but failed to portray the full extent of belief heterogeneity about the duration and impact of the conflict. The gap between synthetic and realized uncertainty  reflects a clean separation between \textit{communication-induced} uncertainty, which the model captures, and \textit{event-driven} uncertainty originating outside the transcript.

All in all, this test demonstrates that our LLM-based approach does not memorize history but rather interprets it, extracting transferable linguistic features that capture genuine policy uncertainty in conversations it has never encountered.\footnote{We also verified that our Gemini API configuration does not enable external grounding or retrieval tools. Specifically, the \texttt{google\_search\_retrieval} 
parameter is disabled, and API response metadata confirm the absence of 
\texttt{groundingMetadata} fields. This ensures the model operates solely 
from its training corpus and cannot access real-time information about 
post-cutoff conferences.}

\subsection{Endogeneity of Agent-Panel Composition}
\label{sec:endogeneity}

Another concern with the agent-based simulation is within-call confounding.
Because panel generation and rate forecasting occur within the same prompt in
the baseline design, the LLM may implicitly condition the characteristics of
the 30 synthetic traders on the transcript it has already processed. If so,
$\widehat{\sigma}_t$ would not be a pure measure of disagreement arising from
heterogeneous processing of a common signal: part of its variation would instead
reflect the LLM's implicit assessment of transcript complexity, encoded directly
into panel composition rather than into forecasts. The correlation between
$\widehat{\sigma}_t$ and realised OIS volatility would persist, but its
interpretation would be fundamentally weakened: the LLM would be detecting
ambiguity and manufacturing dispersion top-down through panel construction,
rather than allowing disagreement to emerge organically from the interaction of
fixed trader characteristics with transcript content. This is not an endogeneity
problem in the classical econometric sense, no regression coefficient is being
biased, but a construct validity concern: the measure may not identify the
behavioral mechanism it purports to capture. We address it through a two-stage
experiment that separates panel generation from forecasting.

In \textit{Stage~1}, the LLM constructs the panel of 30 synthetic traders using
only the macroeconomic regime prevailing around conference date~$t$, with an
explicit instruction not to condition on any press conference content (Figure \ref{fig:panel-prompt}). In
\textit{Stage~2}, each frozen panel is re-presented to the LLM together with
the actual transcript, and individual rate forecasts are elicited exactly as in
the baseline. Crucially, the panel cannot change between stages: whatever
heterogeneity the LLM encodes into trader characteristics in Stage~1 is fixed
prior to transcript exposure, so any dispersion in Stage~2 forecasts must arise
from heterogeneous processing of a common signal rather than from endogenous
panel composition (Figure \ref{fig:forecast-prompt}).\footnote{The simulation has two sources of randomness: which synthetic traders  are selected (Stage~1) and what forecasts they produce (Stage~2). To capture 
both, we draw a fully independent panel for each of the $R = 5$ runs rather 
than fixing the panel across runs. The disagreement measure 
$\widehat{\sigma}^{\text{endo}}_{t}$ is the average within-run standard 
deviation of trader forecasts across the five runs. We use $R = 5$ rather than 
$R = 10$ to limit computational cost.}

We apply the test to a stratified random sample of 30 conferences, drawing 10
per tercile of realised post-meeting OIS volatility, so that the sample spans
low-, medium-, and high-volatility episodes equally.

\begin{figure}[H]
  \centering
  \caption{Stage~1 Panel Composition over Time by Archetype Share}
  \label{fig:endo_composition}
  \includegraphics[width=\linewidth]{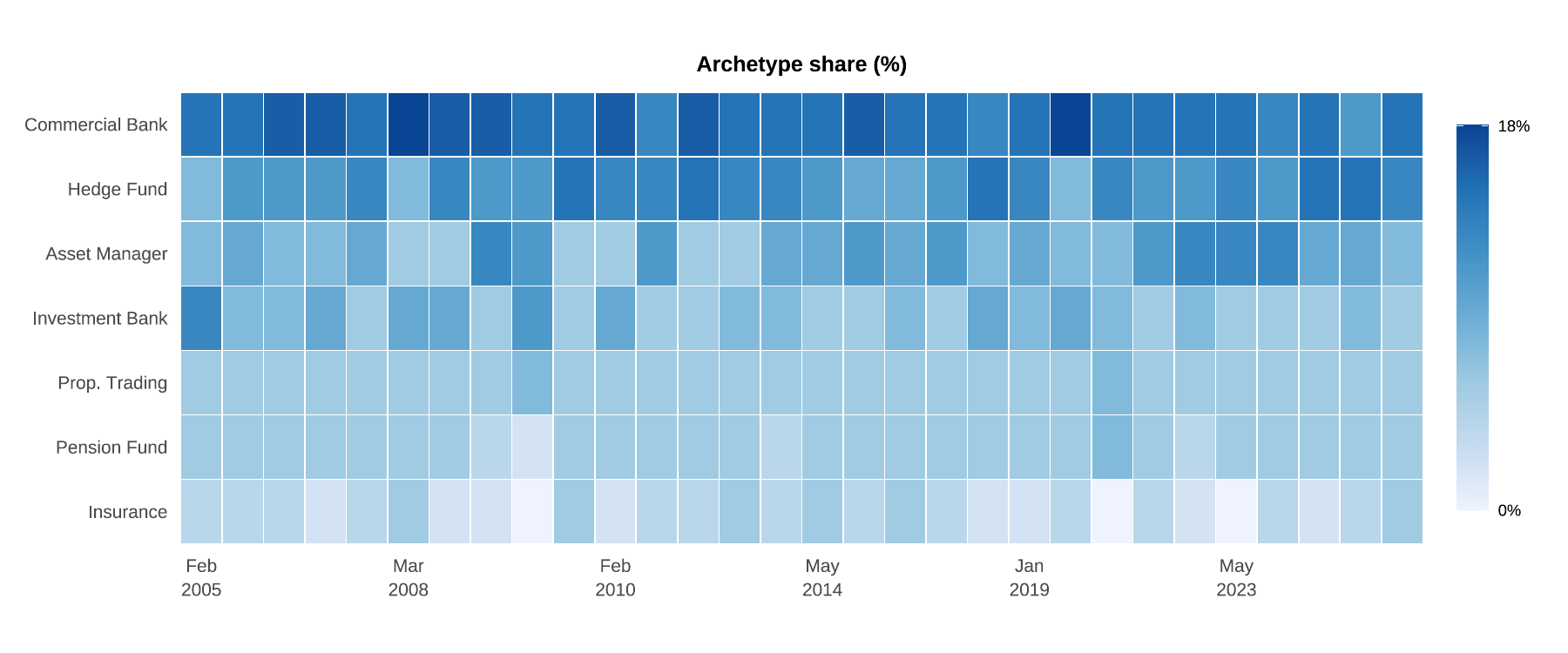}
\end{figure}

Figure~\ref{fig:endo_composition} provides a preliminary validation that the Stage~1 panels encode economically meaningful structure rather than reflecting arbitrary classification.

The archetype heatmap indicates that overall composition is broadly stable over time, with Commercial Banks remaining the dominant archetype throughout the sample, consistent with evidence that euro OIS markets have remained bank‑centred even as their microstructure evolved \citep{ECB2020OIS}. At the same time, the figure exhibits pronounced cyclical variation that aligns closely with monetary policy regimes. In particular, Hedge Fund representation increases during tightening episodes and contracts markedly during the zero lower bound period, in line with the documented decline in speculative and relative‑value interest rate trading under zero‑interest‑rate environments and its resurgence following monetary tightening \citep{EhlersTodorov2025}.\footnote{As a further check, we re-ran the baseline with a five-year stylized description of the composition of the euro OIS market provided directly in the prompt. Results were virtually unchanged, consistent with the LLM being capable of replicating the relevant market structure autonomously, without requiring external guidance on archetype shares.} Taken together, these patterns suggest that Stage~1, reassuringly, captures
economically relevant macro-financial dynamics without recourse to transcript
content. We now turn to the formal comparison between the two designs.

\begin{figure}[H]
  \centering
  \caption{Correlation between Baseline and Two-Stage Disagreement}
  \label{fig:endo_scatter}
  \includegraphics[width=\linewidth]{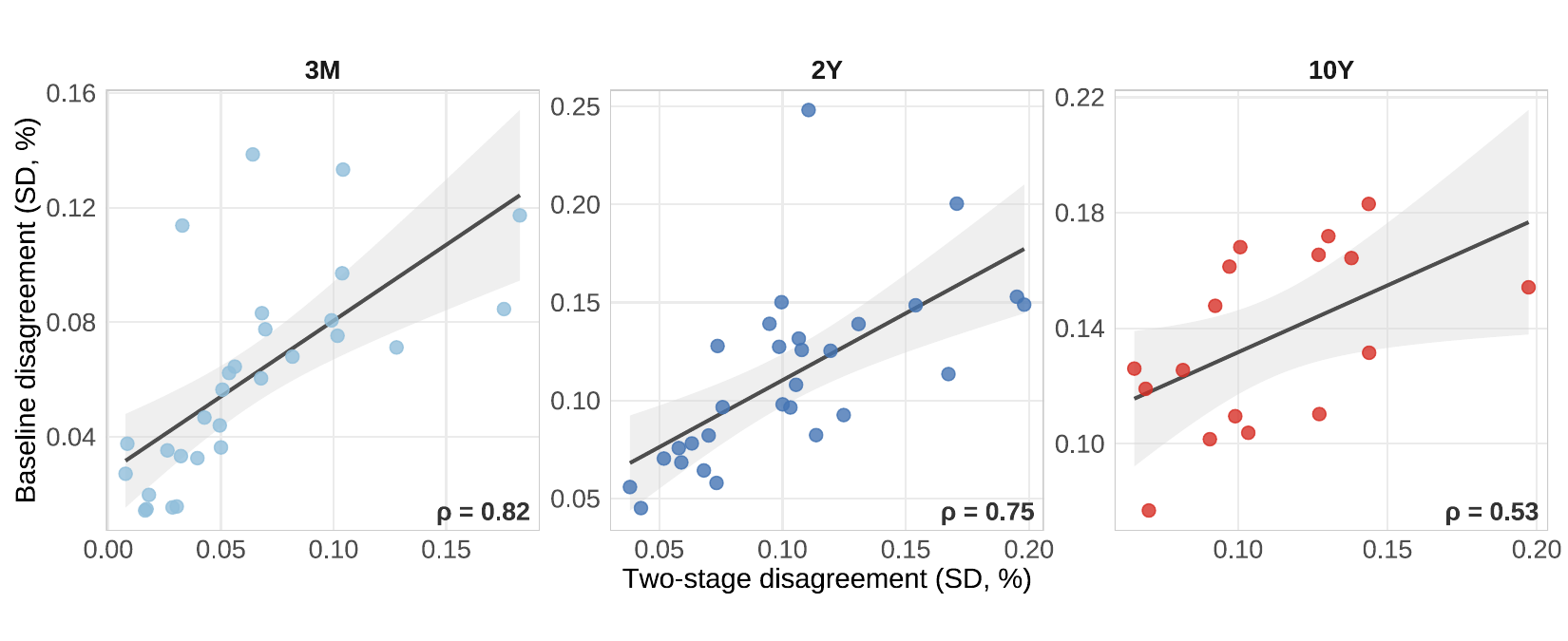}
      \scriptsize\begin{flushleft}
        \textbf{Note:}  Shaded
bands are 95\% confidence interval. Spearman rank correlations $\rho$ are reported in the bottom-right corner of each panel.
\end{flushleft}
\end{figure}

Figure~\ref{fig:endo_scatter} plots the two disagreement measures against each
other. Spearman rank correlations between the baseline and two-stage measures
are 0.82 at 3M, 0.75 at 2Y, and 0.53 at 10Y,
indicating broadly similar rank orderings across both designs at every tenor.

Moreover, across all three tenors, the difference in Spearman correlations between the
baseline and two-stage designs is small in magnitude and statistically indistinguishable from zero.\footnote{95\% bootstrap percentile
confidence intervals: $[-0.155, 0.185]$ at 3M; $[-0.143, 0.352]$ at 2Y;
$[-0.480, 0.431]$ at 10Y; $R = 5{,}000$ replications. The wider interval at
10Y reflects the smaller matched sample ($n = 17$) due to gaps in the baseline
ensemble at that tenor; results there should be interpreted with corresponding
caution.} We therefore conclude that the
correlation between synthetic disagreement and realised OIS volatility
documented in Section~\ref{sec:results} is unlikely to be driven by artefacts of joint panel-and-forecast generation, while acknowledging that the
modest robustness sample limits the precision with which any residual
endogeneity can be bounded.

\subsection{Volatility Persistence Test}

A final question worth exploring is whether our measure merely reflects volatility clustering, an autoregressive phenomenon common in financial markets and unrelated to textual content. To address this, we estimate multivariate regressions of post-announcement disagreement on both pre-conference volatility and our LLM-based measure:

\begin{equation}
\text{Post-Vol}_{t,m} = \alpha + \beta_1 \text{LLM-Std}_{t,m} + \beta_2 \text{Pre-Vol}_{t,m} + \epsilon_{t,m}
\end{equation}

where $t$ indexes press conferences and $m$ denotes maturity.

\input{tables/robustness/p1_pre_post_regression}

The results (Table 2) reveal that both channels operate simultaneously. Pre-conference volatility alone explains 24.8\% confirming that volatility persistence is present in our setting. When adding the LLM-based measure, explanatory power rises to about 38\%. Crucially, both coefficients remain highly significant in the joint specification, indicating that textual content contains predictive information beyond what recent market conditions would imply. Interestingly, the pre-conference volatility coefficient falls by 25\% when controlling for LLM disagreement (from 0.433 to 0.323). This suggests that the LLM measure partially mediates the persistence channel.

Column (3) further includes maturity fixed effects to account for the systematically different unconditional volatility levels across the 3-month, 2-year, and 10-year tenors. Explanatory power rises modestly to 40.3\%, while the coefficient on LLM disagreement remains virtually unchanged at 0.314, confirming that the textual signal is not driven by tenor composition.

These findings clarify the nature of our contribution. For pure forecasting 
of post-announcement volatility, a simple autoregressive benchmark provides 
a useful baseline. However, the LLM framework offers a capability that lagged volatility cannot: it can be applied to draft text \textit{before} 
publication. Developing interpretable methods that identify which specific textual features drive disagreement remains an important direction for future work.

\pagebreak

\section{Conclusions}\label{conclusions}

We show that LLMs can simulate heterogeneous market reactions to central bank communication with economically significant accuracy. Analyzing 293 ECB press conferences spanning the period 1998–2026, we show that cross-sectional dispersion among 30 synthetic traders, each endowed with distinct risk preferences and behavioral biases, correlates at approximately 0.5 with realized post-conference OIS volatility. This performance substantially exceeds textual benchmarks (correlations of 0.1–0.2), adds information beyond volatility clustering, and remains robust in out-of-sample validation using genuinely unseen conferences from January 2025 onwards.

The framework offers practical value across multiple domains. For central banks, it provides an operational tool to anticipate communication-induced volatility before release, transforming communication strategy from reactive refinement to proactive optimization. For researchers, it establishes a micro-founded approach to studying expectation formation that explicitly models heterogeneous belief processes rather than treating them as a black box. Beyond the ECB, the methodology transfers naturally to other major central banks, enabling comparative analysis of how communication styles and institutional frameworks affect interpretive disagreement. More broadly, our results demonstrate that LLMs can serve as computational representations of human interpretive processes, in line with \cite{horton2023}, opening new avenues for modeling belief formation in economics.

Several limitations suggest directions for future research. First, our out-of-sample validation, though encouraging, is necessarily limited to ten post-training conferences. As additional data become available, more comprehensive testing across diverse policy regimes will be essential. Second, domain-specific fine-tuning on central bank communications in the spirit of MILA \citep{bundesbank2024mila} could enhance performance beyond our general-purpose models. Third, the framework could be extended to a full ABM-LLM setting in which agents trade among themselves and new equilibria emerge dynamically (e.g. \citet{delriochanona2025generativeaiagentsbehave}). Finally, theoretical extensions could formalize the strategic dimension of communication design. While our framework enables central banks to minimize interpretive disagreement, optimal communication may require balancing clarity against other objectives. 

To conclude, this work establishes a systematic framework for quantifying communication-induced market disagreement before release. By modeling the heterogeneous interpretive processes through which policy signals transmit to asset prices, we move beyond reduced-form event studies to provide both a diagnostic tool for practitioners and a micro-founded approach for studying expectation formation in monetary policy transmission.

\clearpage
\bibliographystyle{aer}

\input{main.bbl}
\pagebreak



\clearpage
\newpage

\section*{Appendix A: Additional Sensitivity Checks}
\label{app:robustness}

\subsection{Different Correlation Measures}
\label{subsec:correlation}
To ensure the robustness of our main findings, we conduct a supplementary analysis verifying that the observed relationship between synthetic and market-based disagreement is not dependent on the choice of the correlation metric. Our baseline analysis employs the \textbf{Spearman rank correlation coefficient ($\rho$)}, which is well-suited for this context as it captures monotonic relationships and is robust to outliers and non-linearities.

For a more comprehensive validation, we also consider two additional measures: the \textbf{Pearson linear correlation coefficient ($r$)} and \textbf{Kendall's rank correlation coefficient ($\tau$)}:
\begin{itemize}
\item \textbf{Pearson’s $r$} quantifies the strength of a strictly \textit{linear} relationship between two variables. Although more sensitive to outliers, it allows us to check whether the relationship could plausibly be explained within a linear framework.
\item \textbf{Kendall’s $\tau$} provides another non-parametric, rank-based alternative that evaluates concordance between two rankings. It is often preferred in smaller samples or in the presence of tied ranks, offering an additional robustness check alongside Spearman’s $\rho$.
\end{itemize}

\begin{figure}[H]
\centering
\caption{Comparison of Correlation Measures Between Synthetic and Market-Based Disagreement.}
\label{fig:corr_measures}
\includegraphics[width=0.9\textwidth]{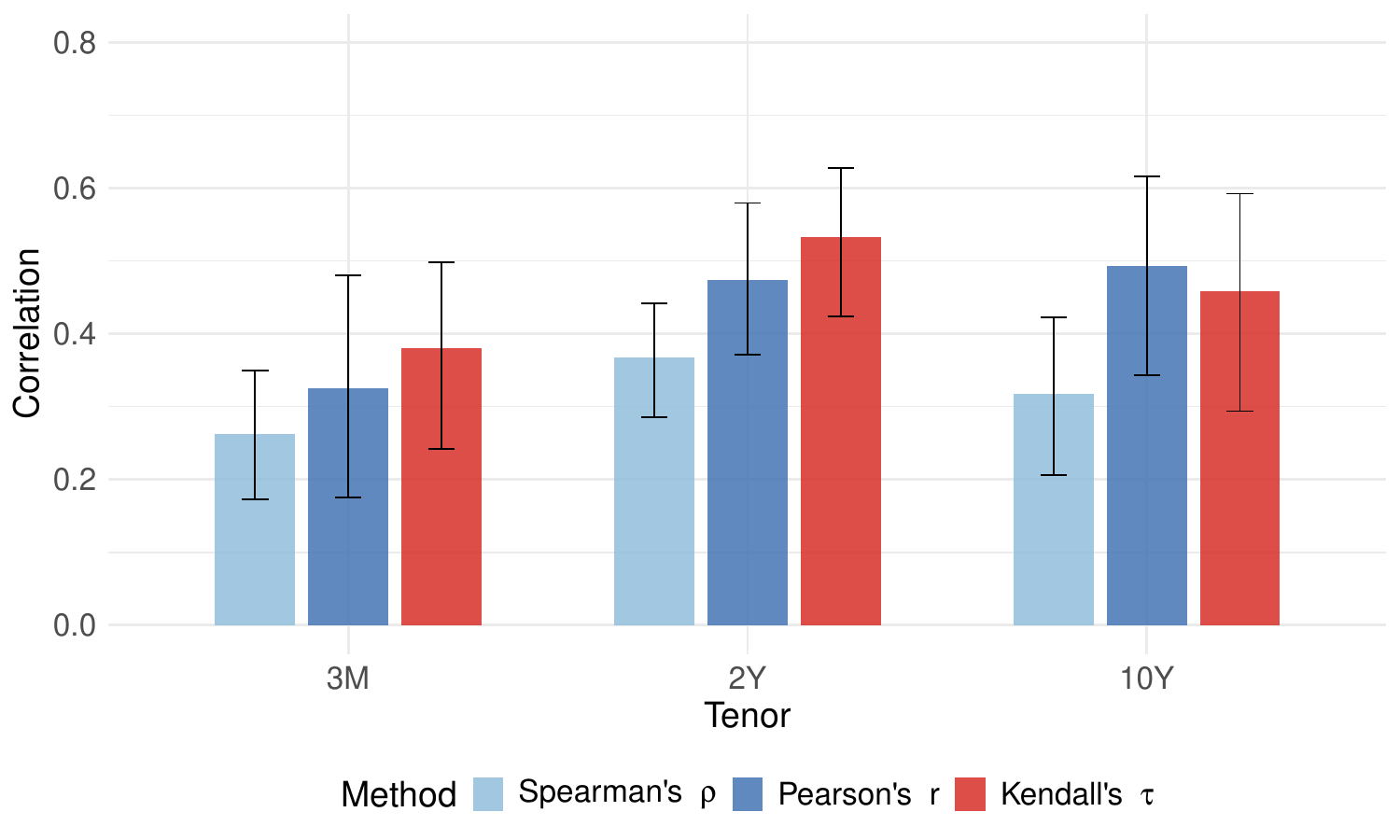}\\
\begin{flushleft}
\scriptsize \textbf{Note}: Correlation between between the output of an LLM model prompted in a zero-shot setting using Gemini 2.5-Flash, with a
temperature parameter of 1, and the OIS min-max range 1 day after ECB conferences. The simulation uses a sample of
293 ECB press conferences (June 1998–March 2026) and the baseline prompt (Figure \ref{fig:llm-prompt}). Correlation measured using Spearman's rank correlation ($\rho$), Pearson's linear correlation ($r$), and Kendall's rank correlation ($\tau$). 95\% confidence intervals are computed via a non-parametric bootstrap with 5,000 replications.
\end{flushleft}
\end{figure}

To assess the precision of these estimates, we compute \textbf{95\% confidence intervals} via a non-parametric bootstrap with 5,000 replications. Figure \ref{fig:corr_measures} presents the results.

Across all three metrics, we find a statistically and economically significant positive association between our LLM-generated disagreement measure and market volatility. The 2-year tenor exhibits the strongest association in the rank-based measures, roughly 0.5, highlighting the simulation’s effectiveness at capturing uncertainty around the medium-term monetary policy path.

Importantly, the bootstrapped 95\% confidence intervals for all coefficients across all tenors comfortably exclude zero, underscoring statistical significance. While the magnitudes vary across correlation metrics, consistent with their differing assumptions, the overall conclusion remains unchanged. The alignment of results across linear and rank-based approaches demonstrates that the link we identify is a robust and stable feature of the data.

\subsection{Prompt and Model Stability Analysis}
\label{subsec:invariance}

Beyond sampling stochasticity, two further implementation choices could in principle affect our disagreement measure: the specific prompt $p$ used to elicit forecasts, and the underlying LLM $m$. We assess robustness along both dimensions using the same reliability framework as in Section~5.1, borrowing from \cite{clayton2025}.

\subsubsection*{Framework}

For each conference $t$, the disagreement measure depends on the prompt $p$,
the model $m$, and a sampling
realisation $s$. We model the resulting outcome as
\begin{equation}
\sigma_t(p, m, s) \;=\; \bar\eta \,+\, \eta_t \,+\, \varepsilon_t(p, m, s),
\label{eq:dgp}
\end{equation}
where $\eta_t$ is the conference-level signal, the across-conference
variation of interest, and $\varepsilon_t(p, m, s)$ is the noise introduced
as we span the set of admissible equivalent prompts $\tilde{\mathcal{P}}$,
vary the model across a set $\tilde{\mathcal{M}}$, and draw fresh sampling
realisations. Define
\[
\sigma^{2}_{\text{between}} \;=\; \Var(\eta_t),
\qquad
\sigma^{2}_{d} \;=\; \Var\bigl(\varepsilon_t \mid d\bigr),
\quad d \in \{\text{sampling}, \text{prompt}, \text{model}\},
\]
where $\sigma^{2}_{d}$ is the variance of $\varepsilon_t$ along dimension $d$
holding the other two fixed at their baseline values. The reliability of a
single measurement of $\sigma_t$ along dimension $d$ is
\begin{equation}
G_d \;=\; \frac{\sigma^{2}_{\text{between}}}
                {\sigma^{2}_{\text{between}} + \sigma^{2}_{d}},
\label{eq:G_d}
\end{equation}
which is the share of variance in a single draw of $\sigma_t$ attributable to
genuine cross-conference variation rather than to noise along dimension $d$.

\subsubsection*{Estimation}
For each dimension $d \in \{\text{prom}, \text{mod}\}$, let $\mathcal{D}_d$ 
denote the set of perturbations (size $K=15$ for prompts, $M=3$ for models), 
and let $\bar\sigma_t^{(d)} = |\mathcal{D}_d|^{-1}\sum_{\delta \in \mathcal{D}_d} 
\sigma_t(\delta)$ be the conference-level mean across perturbations. The 
between-conference variance is estimated as
\begin{equation}
\hat\sigma^{2}_{\text{between},d} \;=\; \Var_t\bigl(\bar\sigma_t^{(d)}\bigr),
\label{eq:between}
\end{equation}
and the within-conference noise variance as
\begin{equation}
\hat\sigma^{2}_{d} \;=\; \frac{1}{T}\sum_{t=1}^{T}
\Var_{\delta \in \mathcal{D}_d}\bigl(\sigma_t(\delta)\bigr),
\label{eq:within}
\end{equation}
where $T = 293$. The G-coefficient is then 

\begin{equation}
\hat{G}_d = \hat\sigma^{2}_{\text{between},d} 
/ (\hat\sigma^{2}_{\text{between},d} + \hat\sigma^{2}_{d})
\end{equation}

Since prompt and model perturbations are each evaluated at a single draw 
($R=1$) due to computational cost constraints, the noise variance 
$\sigma^{2}_{d}$ absorbs residual sampling stochasticity in addition to 
true implementation sensitivity; the corresponding G-coefficients therefore 
represent approximate lower bounds on reliability rather than exact variance 
decompositions.

\subsubsection{Prompt Stability}
\label{subsec:prompt_robustness}

LLMs are particularly sensitive to prompt variations, with ``performance differences of up to 76 percentage points for subtle formatting changes'' \citep{sclar2024quantifyinglanguagemodelssensitivity}. To assess this vulnerability in our context, we generate 15 prompt perturbations of our baseline specification, including 10 minor and 5 medium variations. Minor variations substitute semantically equivalent alternatives and reorder prompt components, while medium variations employ substantively different language to describe trader characteristics and market conditions while maintaining the overall framework. We report the full list of variations in Table 4. 

\input{tables/robustness/prompt_variations_table}

\begin{figure}[H]
    \centering
    \caption{Reliability across Prompt Variations by Tenor}
    \label{fig:icc_by_tenor}
    \includegraphics[width=0.6\textwidth]{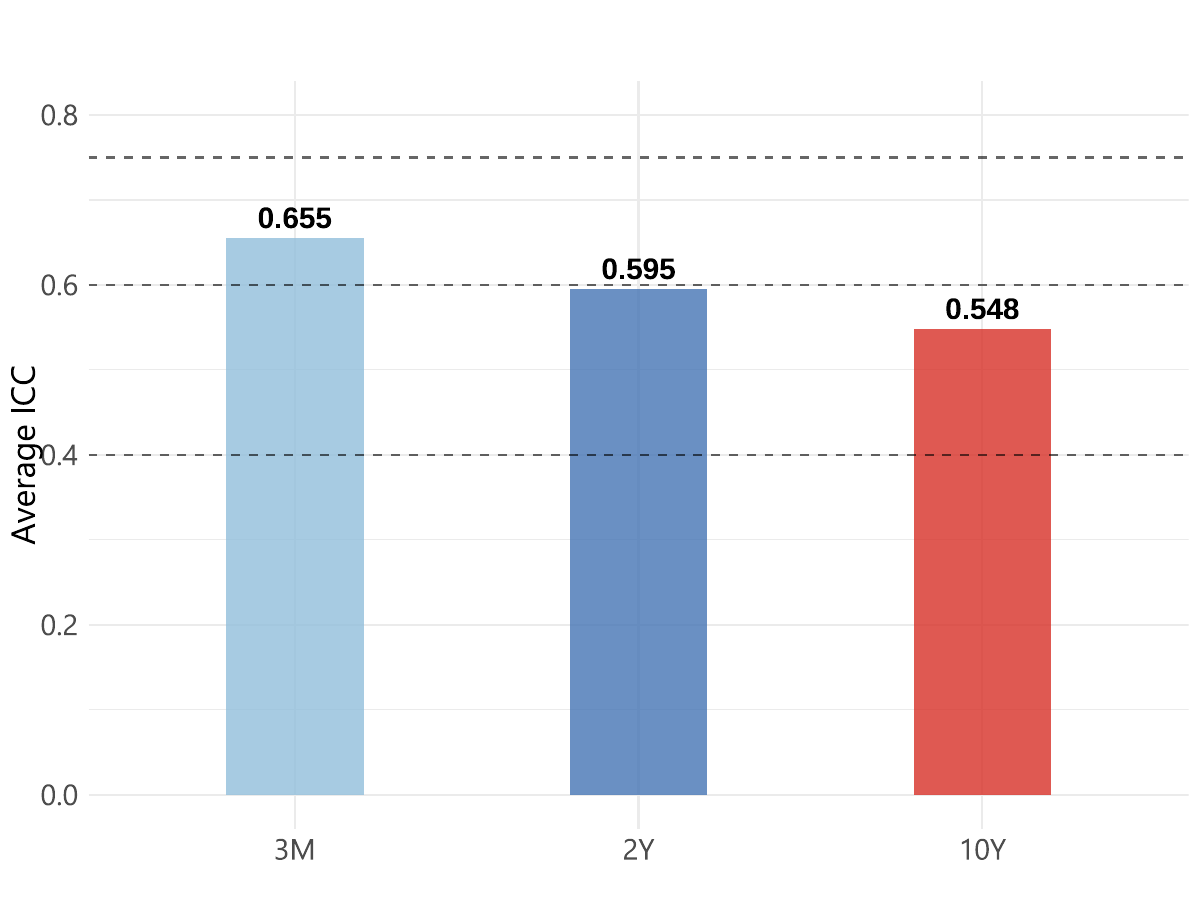}
    \scriptsize{\begin{flushleft}
    \textbf{Note:} Average reliability values across all prompt variations (10 minor and 5 medium) for each tenor. Higher reliability indicates greater measurement stability. Variations are based on prompt \ref{fig:llm-prompt} and obtained by calling Gemini 2.5-Flash. Dashed horizontal lines indicate 0.4 (fair), 0.6 (good), and 0.75 (excellent) thresholds based on \cite{cicchetti1994}.
    \end{flushleft}}
\end{figure}

Given the large computational resources required to run all perturbations on all conferences (283 conferences × 15 perturbations = 4,245 API calls), we select a random subsample of 30 conferences to balance statistical power with computational feasibility. Figure \ref{fig:icc_by_tenor} shows the results.

Reliability follows a clear maturity hierarchy: the 3-month tenor exhibits the highest overall reliability (average ICC = 0.655), followed by the 2-year (ICC = 0.595) and 10-year (ICC = 0.548) tenors (Figure \ref{fig:icc_by_tenor}).  The perturbation analysis yields reassuring results across all 
tenor-specification combinations. While reliability values do not uniformly reach 
excellent standards, they are best interpreted as conservative lower bounds: 
since each perturbation is evaluated at a single draw ($R=1$), sampling noise 
inflates greatly the estimated noise variance and mechanically depresses the 
G-coefficient. The true prompt reliability plausibly lies above these 
estimates, suggesting that conference-specific variation is the dominant 
source of variation in our measure.

\subsubsection{Model Stability}
\label{subsec:model_robustness}

We assess reliability across three distinct LLM architectures: Gemini, ChatGPT, and Claude.\footnote{Specifically, we use Gemini 2.5-Flash, ChatGPT-5o-mini, and Claude-4.5-Sonnet.} These models were selected to capture diversity in training data, design philosophy, and reasoning capabilities. For each ECB conference, we generate synthetic disagreement measures, compute correlations with market-based values, and evaluate cross-model agreement.

\begin{figure}[H]
    \centering
        \caption{Reliability across LLMs by Tenor and Ensemble Model Correlation with Market-based disagreement}
    \label{fig:rob_model_perturbation}
    \includegraphics[width=0.9\textwidth]{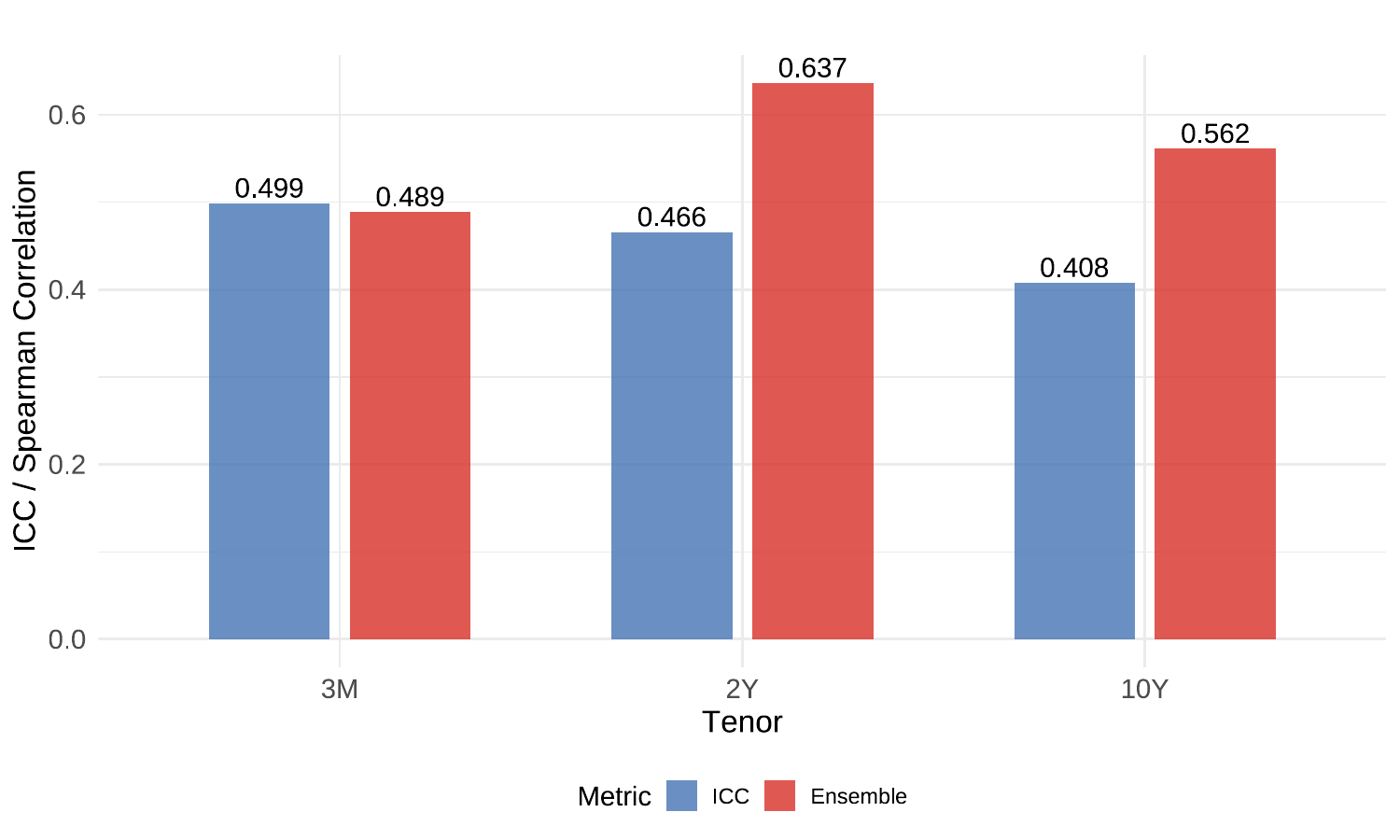}
    \scriptsize{\begin{flushleft}
\textbf{Note:} Blue bars show reliability for each tenor calculated using three distinct LLM architectures (Gemini 2.5 Flash, ChatGPT 5o-mini, Claude 4.5 Sonnet). Higher reliability indicates greater measurement stability. Red bars show the Spearman correlation between the ensemble of LLM outputs, a simple average, and market-based disagreement measured as the min-max range of OIS rates 1 day after ECB press conferences.
    \end{flushleft}}
\end{figure}

The results show fair agreement among models and the same tenor hierarchy displayed in the prompt perturbation exercise: 3-month tenor (ICC = 0.499), 2-year tenor (ICC = 0.466), and 10-year tenor (ICC = 0.408). These ICC values fall in the moderate range, indicating that model-specific differences, however, contribute meaningfully to the observed variance. This raises an important question: do these differences reflect measurement noise, or do different models capture complementary signal?

\begin{figure}[h!]
\centering
\caption{Individual Model Correlations with Market-Based Disagreement by Tenor}
\label{fig:individual_model_correlations}
\includegraphics[width=0.9\textwidth]{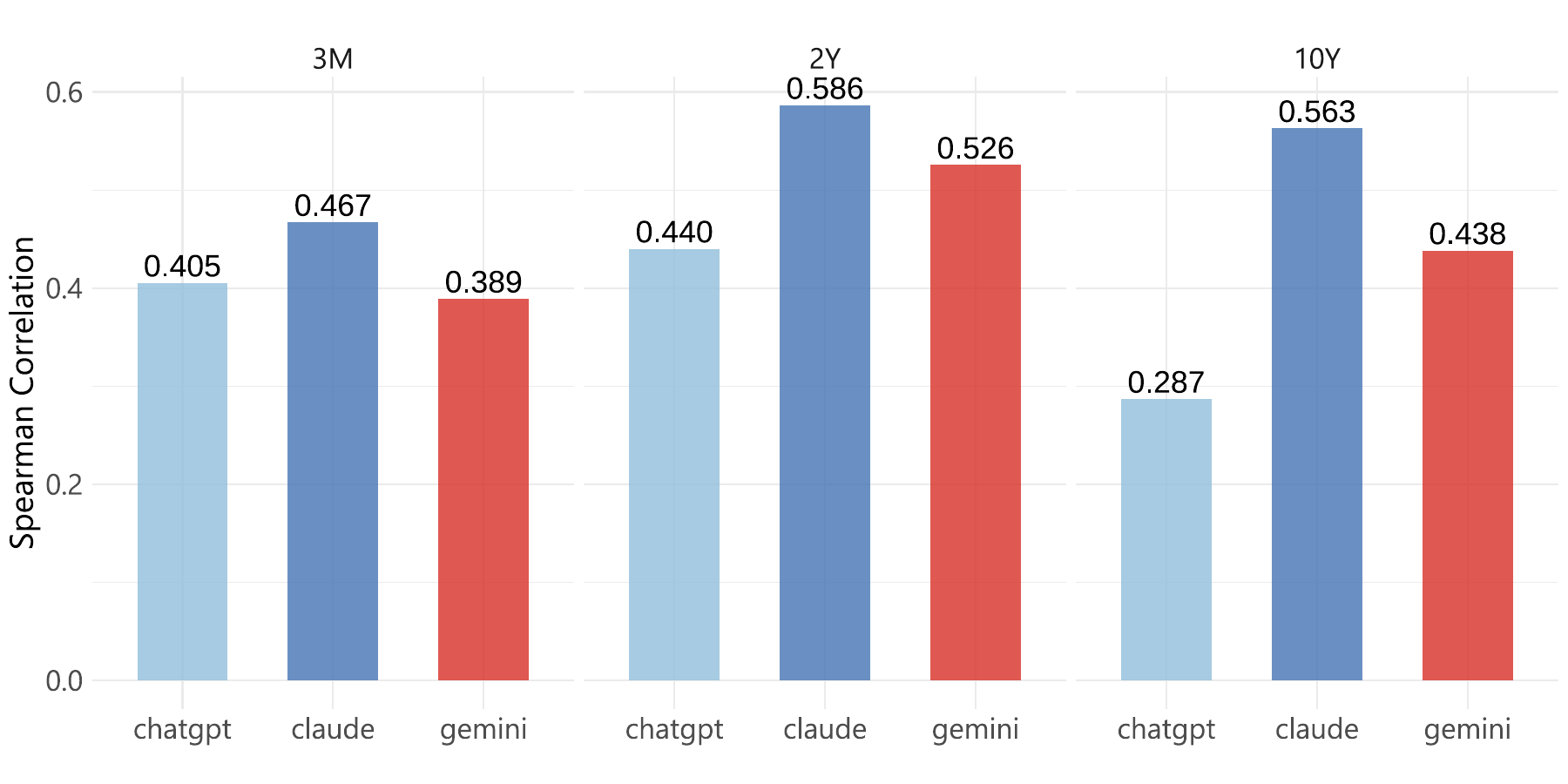}\\
\scriptsize \begin{flushleft} \textbf{Note}: Spearman correlation coefficients between synthetic disagreement measures generated by individual LLM models (Gemini 2.5-Flash, ChatGPT 5o-mini, and Claude Sonnet 4.5) and post-conference OIS volatility across three tenors (3-month, 2-year, and 10-year). Each model generates synthetic disagreement by simulating 30 heterogeneous traders interpreting ECB press conference transcripts. Post-conference market volatility is measured as the min-max range of OIS rates 1 day after ECB press conferences. Sample covers 283 ECB Governing Council meetings from June 1998 to April 2025.
\end{flushleft}
\end{figure}

Two pieces of evidence support the latter interpretation. First, each individual model achieves substantial correlations with market-based disagreement, approximately 0.5 across all tenors (Figure \ref{fig:individual_model_correlations}), demonstrating that all three architectures successfully capture the relationship between textual features and market reactions, albeit through different pathways. Second, and more tellingly, an ensemble measure constructed as the simple average of all three models' outputs achieves higher correlations with market disagreement than any individual model (Figure \ref{fig:rob_model_perturbation}). This performance gain from ensembling is inconsistent with pure noise; if model differences reflected only measurement error, averaging would not improve predictive accuracy.

These findings support a specific interpretation: while individual LLMs are not perfectly interchangeable, their systematic differences capture distinct but relevant aspects of communication interpretation. Different models may weight linguistic features differently, syntactic structure versus semantic content, explicit statements versus implied meanings, yet all extract genuine signal from ECB communication.

\section*{Appendix B: LLM-as-Judge Implementation Details and Results}\label{sec:Appendix}

\begin{algorithm}[H]
\caption{LLM-as-a-Judge Prompt Optimization Algorithm}
\label{alg:llm_optimization}
\begin{algorithmic}[1]
\tiny
\Require Set of $N$ historical transcripts $\mathcal{T} = \{T_1, T_2, \dots, T_N\}$
\Require Initial Analyst LLM prompt $P_{\text{analyst},0}$
\Require Judge LLM prompt $P_{\text{judge}}$
\Require Number of optimization iterations $M$
\Require Correlation threshold $C_{\text{threshold}}$
\Require Analyst LLM $A$, Judge LLM $J$
\Require Number of agents $M$
\Ensure Optimized Analyst LLM prompt $P_{\text{analyst}}^*$

\State Split $\mathcal{T}$ into training set $\mathcal{T}_{\text{train}}$ and test set $\mathcal{T}_{\text{test}}$
\State Initialize optimization history $\mathcal{H} \gets \emptyset$
\State Initialize $P_{\text{analyst,current}} \gets P_{\text{analyst},0}$
\State Initialize $P_{\text{analyst}}^* \gets P_{\text{analyst},0}$
\State Initialize $C_{\text{max}} \gets -\infty$

\For{$m = 1$ to $M$}
    \Comment{Phase 1: Simulate Rate Predictions}
    \State $\mathcal{D}_{\text{sim}} \gets \emptyset$
    \ForAll{$T_i \in \mathcal{T}_{\text{train}}$ \textbf{in parallel}}
        \State Construct user message for Analyst LLM using $T_i$
        \State $D_{\text{sim},i} \gets A(\text{model} = \text{Gemini 2.5 Flash}, \text{prompt} = P_{\text{analyst,current}}, \text{temperature} = 1)$
        \State Parse $D_{\text{sim},i}$ into structured format
        \State Add $D_{\text{sim},i}$ to $\mathcal{D}_{\text{sim}}$
    \EndFor
    \State Compute average correlation of simulated standard deviation with market-based

    \Comment{Phase 2: Judge LLM Evaluation and Prompt Update}
    \State Construct Judge LLM input with:
    \State \quad - Current prompt $P_{\text{analyst,current}}$
    \State \quad - Current correlation between simulated and market-based volatility
    \State \quad - Optimization history $\mathcal{H}$
    \State $O_{\text{judge}} \gets J(\text{model} = \text{Gemini 2.5 Pro}, \text{prompt} = P_{\text{judge}}, \text{temperature} = 0.0)$
    \State Parse $O_{\text{judge}}$ to extract new prompt $P_{\text{analyst,next}}$ and rationale
    \State Append $\{P_{\text{analyst,current}}, \text{avg\_correlation}, O_{\text{judge}}\}$ to $\mathcal{H}$
    \State $P_{\text{analyst,current}} \gets P_{\text{analyst,next}}$
    \State $C_{\text{current}} \gets \text{avg\_correlation}(P_{\text{analyst,current}})$

    \If{$C_{\text{current}} > C_{\text{max}}$}
        \State $C_{\text{max}} \gets C_{\text{current}}$
        \State $P_{\text{analyst}}^* \gets P_{\text{analyst,current}}$
    \EndIf

    \If{$C_{\text{max}} \geq C_{\text{threshold}}$}
        \State \textbf{break}
    \EndIf
\EndFor

\Comment{Phase 3: Final Evaluation}
\State Evaluate $P_{\text{analyst}}^*$ on $\mathcal{T}_{\text{test}}$ using Analyst LLM
\State Compute final performance metrics (correlation)

\Return $P_{\text{analyst}}^*$
\end{algorithmic}
\end{algorithm}

\clearpage
\newpage

\subsection{LLM-as-a-Judge Results}
\label{sec:judge}

The ``Judge'' framework introduces an explicit feedback loop into the simulation. In this setting, a separate LLM evaluates the forecasting performance of the simulated agents and rewrites the prompt to improve alignment with observed market reactions. This design allows the forecasting agents to adapt iteratively: while the baseline setup relies on fixed prompts, the Judge framework endogenizes prompt selection, allowing us to evaluate whether prompt optimization alone can account for improvements in performance.  We continue this back-and-forth for a total of 5 iterations, testing the new prompts on a training set, and then use every prompt produced for the test set.\footnote{Each prompt is run 5 times to reduce sampling noise. This results in 50 simulations per conference (5 runs × 5 iterations × 2 datasets: training and test). This total must then be multiplied by the number of ECB press conferences in our sample.} The stopping criterion emerged naturally: at the 6th iteration, the Judge attempted to reframe the task from predicting rate levels to basis point changes, which would have compromised our cross-sectional dispersion measurement framework. We therefore terminated optimization at iteration 5 to preserve methodological consistency.

\begin{figure}[!htbp]
    \centering
    \includegraphics[width=0.79\textwidth]{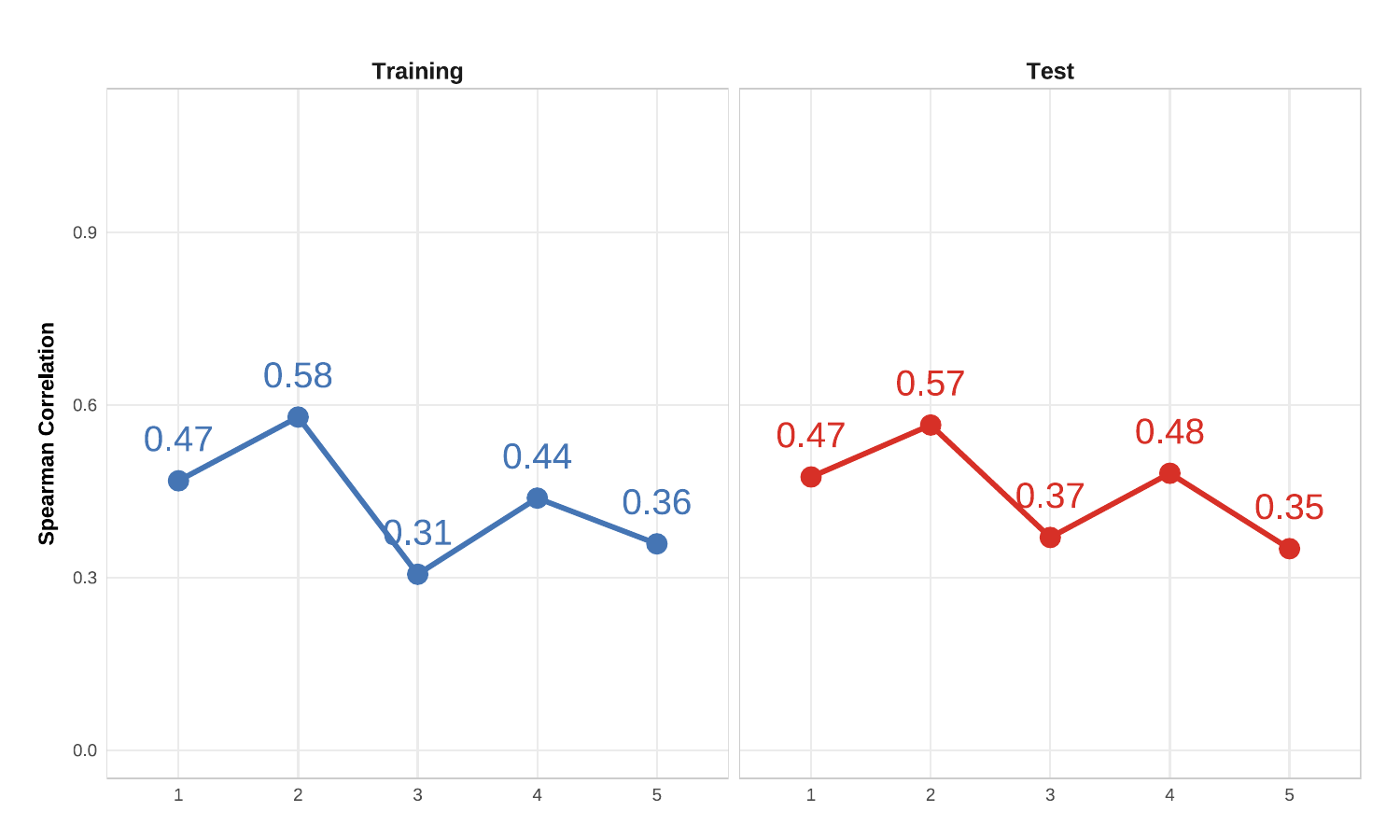}
    \caption{Correlation with market-based measures across iterations --- training and test set}
    \label{fig:avg_correlation_judge}
    \scriptsize\begin{flushleft}
    \textbf{Note:} Each point reflects the average Spearman correlation across tenors (3-month, 2-year and 10-year) between LLM-generated standard deviation of Euro OIS rate forecasts and the OIS min-max range 1 day after ECB conferences. The x-axis indicates the prompt iteration in the LLM-as-a-Judge framework. The simulation uses Gemini 2.5-Flash with temperature 1 and incorporates a feedback loop via a separate Judge LLM using Gemini 2.5-Pro. Train and test set are split 75/25 over June 1998--April 2025.
\end{flushleft}
\end{figure}

Figure \ref{fig:avg_correlation_judge} reveals a striking pattern: the Judge's refinements follow a clear performance arc across both training and test sets. Performance peaks dramatically at iteration 2, jumping from 0.47 to 0.58 in training, a substantial 11 percentage points improvement that holds remarkably well out-of-sample (0.57). This consistency across datasets suggests the Judge identified genuine signal rather than noise. Beyond iteration 2, however, the optimization process exhibits declining effectiveness. Iteration 3 marks a sharp deterioration, with correlations dropping to 0.31-0.37, before partially recovering in the 4th iteration (0.44-0.48) and then partially dropping again (0.36-0.35). This rollercoaster pattern reveals a key insight: more sophisticated prompts do not always yield better results. The Judge's initial refinement successfully linked textual ambiguity to prediction dispersion, but subsequent "improvements" appear to have over-engineered the task, constraining the model's natural interpretive flexibility.

The key to the Judge's success in the first iteration lies in a deceptively simple insight: the second prompt achieves peak performance by establishing an explicit mapping between textual properties and prediction dispersion. As the Judge observes, the initial specification lacked ``explicit instruction linking the ambiguity or clarity of the ECB press conference transcript to the dispersion of the traders' predictions" (Table \ref{tab:prompt_judge}). Where the original prompt left this connection implicit, the refined version instructs agents that their collective disagreement "must directly reflect the level of ambiguity, surprise, or conflicting signals within the ECB press conference transcript." This finding echoes in part our earlier work showing that larger monetary surprises generate correspondingly larger uncertainty surprises \citep{collodel2025market}, though,  remarkably, the Judge discovered this relationship independently through iterative refinement. The Judge's diagnosis proves prescient: when given clear instructions to map vague language, contrasting messages, or deviations from expectations into wider prediction spreads, the LLM excels. Subsequent iterations demonstrate how optimization can become counterproductive. Prompt 3 imposes explicit dispersion targets (1-3 bps for clarity, 8-15 bps for ambiguity), while Prompt 5 mandates multi-factor uncertainty scoring. Both achieve lower correlations (0.31-0.37) than simpler specifications, suggesting these constraints inhibit the model's natural interpretive flexibility. Prompt 4 partially recovers (0.44) by relaxing these rules. This finding has direct implications for central banks considering LLM-based communication tools: automated prompt optimization requires human expertise to distinguish genuine improvements from theoretically appealing but empirically counterproductive complexity.

\input{tables/llm_as_judge/judge_llm_prompts}

The optimization trajectory, from breakthrough to degradation within just a few iterations, underscores that human judgment remains indispensable in guiding AI systems toward practical applications.

\section*{Appendix C: Full Prompts}

\begin{figure}[H]
\centering
\caption{Baseline Prompt}
\label{fig:llm-prompt}
\begin{lstlisting}[basicstyle=\ttfamily\scriptsize, breaklines=true, columns=flexible, frame=single, backgroundcolor=\color{gray!5}, linewidth=\textwidth, xleftmargin=0pt, xrightmargin=0pt]
Context:
You are simulating the Euro area interest rate swap market, composed of 30 individual traders.
These traders interpret the ECB Governing Council press conference, which communicates monetary policy decisions, economic assessments, and includes a Q&A session with journalists.
Each trader then makes a trading decision to maximize profit based on their interpretation of the conference and their unique characteristics.
Trader Characteristics:
Each trader has the following attributes:
- Risk Aversion: High / Medium / Low determines sensitivity to uncertainty and preference for stability
- Behavioral Biases (1,2 per trader): e.g., Confirmation Bias, Overconfidence, Anchoring, Herding, Loss Aversion, Recency Bias
- Interpretation Style (1 per trader): e.g., Fundamentalist, Sentiment Reader, Quantitative, Skeptic, Narrative-Driven
Task:
You are given the text of a single ECB press conference.
For each of the 30 traders, simulate their individual trading action in the interest rate swap market across three tenors (3 months, 2 years, 10 years).
For each tenor, the trader must:
    - Provide an expected rate direction: Up / Down / Unchanged (relative to the pre-conference rate)
    - Provide a new expected swap rate (in percent, to two decimal places)
    - Provide a confidence score (0 to 100%) reflecting how strongly the trader believes in his forecast, based on their interpretation of the press conference and their own characteristics
Output:
Provide a table with the following structure for each press conference, trader, and interest rate tenor

| Date       | Trader ID | Tenor   | Expected Direction | New Expected Rate (%) | Confidence (%) |
|------------|-----------|---------|--------------------|-----------------------|----------------|
| YYYY-MM-DD | T001      | 3M      | Up                 | 3.15                  | 65             |
| YYYY-MM-DD | T001      | 2Y      | Down               | 2.85                  | 80             |
| ...        | ...       | ...     | ...                | ...                   | ...            |
Guidelines:
- Use only the information available as of [date].
- Do not aggregate or summarize responses.
- Reflect diversity in interpretation, risk tolerance, and horizon. Rationale must be unique for each trader and can vary across tenors.
- Output only a markdown table with the specified columns, no additional text. Do not use JSON or any other data serialization format.
\end{lstlisting}
\end{figure}

\begin{figure}[H]
\centering
\caption{Historical Anchoring Prompt}
\label{fig:llm-prompt-historical}
\begin{lstlisting}[basicstyle=\ttfamily\scriptsize, breaklines=true, columns=flexible, frame=single, backgroundcolor=\color{gray!5}, linewidth=\textwidth, xleftmargin=0pt, xrightmargin=0pt]
Context:
You are simulating the Euro area interest rate swap market, composed of 30 individual traders.
These traders interpret the ECB Governing Council press conference, which communicates monetary policy decisions, economic assessments, and includes a Q&A session with journalists.
Each trader then makes a trading decision to maximize profit based on their interpretation of the conference and their unique characteristics.

Trader Characteristics:
Each trader has the following attributes:
- Risk Aversion: High / Medium / Low --- determines sensitivity to uncertainty and preference for stability.
- Behavioral Biases (1--2 per trader): e.g., Confirmation Bias, Overconfidence, Anchoring, Herding, Loss Aversion, Recency Bias.
- Interpretation Style (1 per trader): e.g., Fundamentalist, Sentiment Reader, Quantitative, Skeptic, Narrative-Driven.

Task:
You are given a certain number of distinct ECB press conferences.
For each of the 30 traders, simulate their individual trading action in the interest rate swap market across three tenors (3 months, 2 years, 10 years).
For each tenor, the trader must:
   - Provide an expected rate direction: Up / Down / Unchanged
   - Provide a new expected swap rate (in percent, to two decimal places)
   - Provide a confidence level (0-100%) in their decision

Output:
Provide a table with the following structure for each press conference, trader, and interest rate tenor:

| Date       | Trader ID | Tenor   | Expected Direction | New Expected Rate (%)  | Confidence Level (%) |
|------------|-----------|---------|--------------------|------------------------|----------------------|
| YYYY-MM-DD | T001      | 3M      | Up                 | 3.15                   |                      |
| YYYY-MM-DD | T001      | 2Y      | Down               | 2.85                   |                      |
| ...        | ...       | ...     | ...                | ...                    |                      |

Guidelines:
- Use only the information available as of [date].
- To simplify the task, we provide the before and after standard deviation for the previous three ECB press conferences (for each tenor).
- Do not aggregate or summarize responses.
- Reflect diversity in interpretation, risk tolerance, and horizon. Rationale must be unique for each trader and can vary across tenors.
- Output only a markdown table with the specified columns, no additional text. Do not use JSON or any other data serialization format.
- If multiple press conferences are included, clearly distinguish between them using the 'Date' field.
\end{lstlisting}
\end{figure}

\begin{figure}[H]
\centering
\caption{Panel generation prompt (Stage 1)}
\label{fig:panel-prompt}
\begin{lstlisting}[basicstyle=\ttfamily\tiny, breaklines=true, columns=flexible, frame=single, backgroundcolor=\color{gray!5}, linewidth=\textwidth, xleftmargin=0pt, xrightmargin=0pt]
You are constructing a synthetic panel of 30 market participants who would have
been active in euro-area fixed-income and interest rate swap markets around [date].

CRITICAL CONSTRAINT: Generate this panel based ONLY on the macroeconomic regime
prevailing around [date] -- the interest rate environment, the recent trajectory
of ECB monetary policy, and broad market conditions as of [date]. DO NOT
condition on the content of any specific ECB press conference, any speech, or
any text that follows. Generate the panel based on macroeconomic regime alone.

Panel construction:
- Generate exactly 30 participants, identified as T001 through T030.
- Draw from the realistic range of institutions active in euro-area OIS markets:
  commercial banks, investment banks, hedge funds, pension funds, insurance firms,
  proprietary trading desks, and asset managers.
- Each participant's archetype, risk profile, time horizon, and prior view should
  reflect who was active and how they were positioned in the euro-area rate market
  around [date], given the monetary policy cycle at that time.
- Assign each participant:
    (a) risk_profile  : High / Medium / Low
    (b) time_horizon  : Short-term (< 1 week) / Medium-term (1 week to 3 months) /
                        Long-term (> 3 months)
    (c) prior_view    : Hawkish / Neutral / Dovish -- reflecting their rate outlook
                        as of [date], based on regime alone, not any specific event
    (d) key_bias      : one of Confirmation Bias, Overconfidence, Anchoring,
                        Herding, Loss Aversion, Recency Bias, Status Quo Bias,
                        Narrative Fallacy
- Distribute archetypes, risk profiles, prior views, and biases to reflect the
  realistic heterogeneity of euro OIS market participants around [date].
- Ensure meaningful spread: not all participants should share the same prior view
  or the same key bias.

Macroeconomic regime conditioning (use general knowledge as of [date] only):
- Where was the ECB deposit rate relative to its recent history around [date]?
- Was ECB policy in a tightening, easing, or on-hold phase around [date]?
- What was the prevailing level of uncertainty about future rate paths in the
  euro area at that time?
Do NOT reference any specific press conference, statement, or meeting -- broad
macroeconomic regime information only.

Output:
Return ONLY the following markdown table. No preamble, no commentary, and no
explanation before or after the table. The table must have exactly these six
columns in this order.

| agent_id | archetype       | risk_profile | time_horizon  | prior_view | key_bias          |
|----------|-----------------|--------------|---------------|------------|-------------------|
| T001     | Commercial Bank | Medium       | Short-term    | Neutral    | Anchoring         |
| T002     | Hedge Fund      | High         | Short-term    | Hawkish    | Overconfidence    |
| ...      | ...             | ...          | ...           | ...        | ...               |
| T030     | Asset Manager   | Medium       | Medium-term   | Neutral    | Herding           |

Generate all 30 rows. The agent_id column must run sequentially from T001 to T030.
\end{lstlisting}
\end{figure}

\begin{figure}[H]
\centering
\caption{Transcript-conditioned forecast prompt (Stage 2)}
\label{fig:forecast-prompt}
\begin{lstlisting}[basicstyle=\ttfamily\tiny, breaklines=true, columns=flexible, frame=single, backgroundcolor=\color{gray!5}, linewidth=\textwidth, xleftmargin=0pt, xrightmargin=0pt]
You are conducting a two-stage simulation of the euro-area interest rate swap
market. In Stage 1, a panel of 30 market participants was constructed based
solely on the macroeconomic regime prevailing around [date]. That panel is
provided below. In Stage 2 -- your task here -- you will generate individual
rate forecasts for each participant, conditional on the ECB press conference
transcript provided at the end of this prompt.

FIXED PANEL -- use exactly as given; do not regenerate, rename, reorder,
reinterpret, or add participants:
[panel]

Instructions:
- Use ONLY the 30 participants listed in the panel above (T001-T030).
- For each participant, simulate their individual forecast in the euro-area
  interest rate swap market across three tenors: 3 months (3M), 2 years (2Y),
  and 10 years (10Y), conditional on the press conference transcript below.
- Each participant reads and interprets the transcript through the lens of
  their assigned archetype, risk profile, time horizon, prior view, and key
  behavioral bias.
- For each (participant x tenor) combination, provide:
    (a) Expected direction: Up / Down / Unchanged -- relative to the
        pre-conference rate.
    (b) New expected swap rate in percent, to two decimal places.
    (c) Confidence score (0-100) reflecting how strongly the participant
        believes in their forecast, given their characteristics and their
        interpretation of the transcript.

Output:
Return ONLY the following markdown table -- one row per (participant x tenor)
combination, for all 30 participants and all 3 tenors (90 data rows total).
No preamble, no commentary, no explanation before or after the table.

| Date       | Trader ID | Tenor | Expected Direction | New Expected Rate (%) | Confidence (%) |
|------------|-----------|-------|--------------------|-----------------------|----------------|
| YYYY-MM-DD | T001      | 3M    | Up                 | 3.15                  | 65             |
| YYYY-MM-DD | T001      | 2Y    | Down               | 2.85                  | 80             |
| YYYY-MM-DD | T001      | 10Y   | Unchanged          | 2.60                  | 50             |
| ...        | ...       | ...   | ...                | ...                   | ...            |
| YYYY-MM-DD | T030      | 10Y   | Up                 | 2.75                  | 70             |

Guidelines:
- Use only the information available as of [date].
- Each participant's forecast must reflect their unique characteristics from
  the panel -- archetype, risk profile, time horizon, prior view, and key bias.
- Ensure diversity in interpretation: not all participants should forecast the
  same direction or the same rate level.
- Do not aggregate or summarize responses.
- Output ONLY the markdown table -- no JSON or other format.
\end{lstlisting}
\end{figure}

\begin{figure}[H]
\centering
\caption{Judge prompt}
\label{fig:judge-prompt}
\begin{lstlisting}[basicstyle=\ttfamily\scriptsize, breaklines=true, columns=flexible, frame=single, backgroundcolor=\color{gray!5}, linewidth=\textwidth, xleftmargin=0pt, xrightmargin=0pt]
Context:
You are an expert AI system designed to optimize prompts for another AI (the "Analyst LLM").
Your ultimate goal is to refine the Analyst LLM's prompt to improve its ability to replicate the market volatility of OIS rates based on ECB press conference transcripts.
Specifically, you must ensure that the standard deviation of the Analyst LLM's predictions correlates highly with the actual, observed market volatility of the 3-month, 2-years and 10-year OIS rates.
This means a higher standard deviation in the Analyst's predictions should correspond to higher actual market volatility, and vice-versa.
You will be provided with:
- The current Analyst LLM prompt.
- The most recent performance (Spearman correlation coefficient between the Analyst LLM's predicted standard deviations and actual market volatility).
- The historical performance trend, including past critiques and proposed prompt summaries.
Your task is to:
1. Critique the current prompt: Identify specific weaknesses or areas of ambiguity that might directly hinder achieving a high positive correlation. Consider:
   - Clarity and Specificity: Is the Analyst LLM's task unambiguous?
   - Emphasis on Uncertainty: Does the prompt adequately guide the Analyst to reflect internal uncertainty in its prediction spread?
   - Guidance on Nuance: Does it encourage consideration of subtle market signals from the text?
2. Suggest a Revised Prompt: Propose a new version of the Analyst LLM's prompt that directly addresses the identified weaknesses and aims to increase the correlation. Be precise with your suggested changes.
3. Explain your reasoning: Articulate why your proposed revisions are expected to improve the correlation, linking specific prompt changes to anticipated improvements in the Analyst LLM's behavior regarding uncertainty quantification.
Your output must be in JSON format. Do not include any other text outside the JSON.
Example JSON output:
{
  "critique": "The previous prompt was too general regarding how to express uncertainty. It didn't explicitly ask the Analyst LLM to consider multiple viewpoints, which is key for its standard deviation to accurately reflect market volatility. It also lacked emphasis on how ambiguity in the transcript should translate to higher spread.",
  "revised_prompt": "You are a highly analytical financial expert specializing in macroeconomic analysis, with a focus on central bank communication. Your task is to analyze excerpts from ECB press conferences and predict the immediate percentage change in the 10-year Overnight Index Swap (OIS) rate (in basis points). When analyzing each excerpt, **explicitly consider and internalize the potential range of market interpretations**. If the language is ambiguous, vague, or contains conflicting signals, your internal simulation of potential outcomes should broaden. Conversely, clear and unambiguous guidance should lead to a narrower range. Your final prediction should be a single numerical value (e.g., +5, -2, 0) reflecting your best estimate. The *variability* across multiple independent predictions you generate for the same transcript is expected to directly reflect the market's anticipated uncertainty. Example Format: +5",
  "reasoning": "The revised prompt adds explicit instructions to consider the 'range of market interpretations' and directly links 'ambiguous language' to a 'broadened' internal simulation. This should encourage the Analyst LLM to generate a higher standard deviation in its outputs for uncertain transcripts and a lower standard deviation for clear ones."
}
\end{lstlisting}
\end{figure}

\clearpage
\newpage

\section*{Appendix D: Implementation Details}
\label{app:implementation_details}

\begin{table}[h]
\centering
\small
\caption{LLM Model Specifications and Parameters}
\label{tab:llm_specs}
\begin{tabular}{ll}
\hline
\textbf{Parameter} & \textbf{Value} \\
\hline
\multicolumn{2}{l}{\textit{Model Information}} \\ 
Primary Model & Google Gemini 2.5-Flash \\
Judge Model (LLM-as-a-Judge) & Google Gemini 2.5-Pro \\
Robustness Models & Claude Sonnet 4.5, ChatGPT-5o-mini \\
API Version (Gemini) & v1beta \\
API Endpoint (Gemini) & \texttt{/v1beta/models/} \\
Knowledge Cutoff (Gemini) & January 2025 \\
Knowledge Cutoff (Claude) & June 2025 \\
Knowledge Cutoff (ChatGPT) & May 2024 \\[0.2cm]
\multicolumn{2}{l}{\textit{Generation Parameters}} \\ 
\texttt{temperature} & 1.0 \\
\texttt{topK} & 40 \\
\texttt{topP} & 0.95 \\[0.2cm]
\multicolumn{2}{l}{\textit{API Configuration}} \\
Request timeout & 120 seconds \\
Maximum retry attempts & 5 \\
Retry delays & 5, 10, 15, 20, 25 seconds \\
Parallel workers & 5 \\
\hline
\end{tabular}
\end{table}

\begin{table}[h]
\centering
\small
\caption{Data Processing Specifications}
\label{tab:data_processing}
\begin{tabular}{lp{7cm}}
\hline
\textbf{Aspect} & \textbf{Description} \\
\hline
\multicolumn{2}{l}{\textit{ECB Press Conference Transcripts}} \\
Data Source & Official ECB website \\
Coverage & Full transcripts (opening statement + Q\&A) \\
Text Cleaning & Remove non-alphanumeric except \{space, ., ?, -\} \\
Cleaning Function & \texttt{gsub("[{\textasciicircum}[:alnum:] .?-]", "", text)} \\
Truncation & None applied \\
Sample Period & June 9, 1998 -- March 19, 2026 \\
Total Conferences & 283 \\
Avg N. Tokens & 4456\\
25th--75th Percentile Tokens & 3990 -- 5233\\
\hline
\end{tabular}
\begin{flushleft}
    \scriptsize \textbf{Note:} Token counts are estimated using the standard conversion ratio of 0.75 tokens per word (i.e., 1 word $\approx$ 1.33 tokens). This approximation follows common practice for English text and provides a consistent measure of model input length.
\end{flushleft}
\end{table}


\end{document}

%% file: tables/robustness/complexity_correlations.tex
\begin{table}[!htbp]
\centering
\caption{Spearman Correlations Between Text-Based Measures and Market-Based Disagreement by Tenor}
\label{tab:complexity_correlations}
\renewcommand{\arraystretch}{1.35}

\begin{tabular}{lccc}
\hline\hline
\textbf{Measure} & \textbf{3M} & \textbf{2Y} & \textbf{10Y} \\
\hline

\multicolumn{4}{l}{\textbf{A. Complexity Measures}} \\
\quad FK Complexity              & 0.227*** & 0.229*** & -0.000 \\ 
\quad Word Count                 & 0.157**  & 0.151**  & -0.052 \\[0.3em]

\multicolumn{4}{l}{\textbf{B. Framing and Tone Measures}} \\
\quad Hedging Words              & 0.219*** & 0.158**  & 0.188** \\ 
\quad LM Uncertainty             & 0.161**  & 0.149**  & -0.018 \\[0.3em]

\multicolumn{4}{l}{\textbf{C. Stance / Policy Sentiment Measures}} \\
\quad Net Hawkish--Dovish Score  & 0.026    & 0.215*** & 0.135 \\
\hline\hline
\end{tabular}

\vspace{0.6em}

\begin{minipage}{0.98\textwidth}
\scriptsize
\textbf{Note}: This table reports Spearman rank correlations between text-based measures based on 293 ECB press conferences (June 1998-March 2026) and market-based disagreement post-ECB conferences, measured as the min-max range of Euro OIS rates 1 day after the conference, across three interest rate tenors. The measures are grouped into three conceptual buckets. \textbf{(A) Complexity Measures}: FK Complexity refers to the Flesch--Kincaid Grade Level, which captures linguistic complexity based on sentence length and syllables per word; higher values indicate less readable and more technical language. Word Count represents the total length of the communication and proxies for informational load or verbosity. \textbf{(B) Framing and Tone Measures}: Hedging Words measure the frequency of cautious or non-committal expressions (e.g., ``may,'' ``could,'' ``possibly''), which indicate linguistic uncertainty or reduced commitment. LM Uncertainty is based on the Loughran--McDonald Finance dictionary and captures explicit expressions of uncertainty in economic or financial contexts. \textbf{(C) Stance / Policy Sentiment Measures}: The Net Hawkish--Dovish Score reflects the balance of hawkish versus dovish terms and captures the directional policy tone of the communication. Outliers above the 99th percentile are removed. Significance levels: *** $p<0.01$, ** $p<0.05$, * $p<0.10$.
\end{minipage}

\end{table}

%% file: tables/few-shot/fewshot_vs_naive_calibration.tex
\begin{table}[!htbp] \centering 
  \caption{Bias and Mean Absolute Error in Baseline and Historical Anchoring} 
  \label{tab:mae} 
\small 
\begin{tabular}{@{\extracolsep{5pt}} lcccccc} 
\\[-1.8ex]\hline 
\hline \\[-1.8ex] 
Tenor & N & Bias (N) & Bias (FS) & MAE (N) & MAE (FS) & Delta MAE \\ 
\hline \\[-1.8ex] 
3M & $89$ & 0.014\textasteriskcentered \textasteriskcentered \textasteriskcentered  & -0.007\textasteriskcentered \textasteriskcentered  & 0.030 & 0.026 & 0.004 \\ 
2Y & $89$ & 0.020\textasteriskcentered \textasteriskcentered \textasteriskcentered  & -0.018\textasteriskcentered \textasteriskcentered \textasteriskcentered  & 0.044 & 0.036 & 0.009\textasteriskcentered \textasteriskcentered  \\ 
10Y & $53$ & 0.062\textasteriskcentered \textasteriskcentered \textasteriskcentered  & 0.007 & 0.070 & 0.041 & 0.029\textasteriskcentered \textasteriskcentered \textasteriskcentered  \\ 
Pooled & $231$ & 0.028\textasteriskcentered \textasteriskcentered \textasteriskcentered  & -0.008\textasteriskcentered \textasteriskcentered  & 0.045 & 0.033 & 0.012\textasteriskcentered \textasteriskcentered \textasteriskcentered  \\ 
\hline 
\end{tabular}
\begin{tablenotes}
\scriptsize
\item \textbf{Note}: Bias $N$ and Bias ${FS}$ denote the mean difference between the baseline and historically anchored LLM disagreement measure and realized market volatility, respectively. MAE($N$) and MAE(${FS}$) are the corresponding mean absolute errors. $\Delta$ MAE $=$ MAE($N) -$ MAE(${FS}$) $^{*}p<0.10$, $^{**}p<0.05$, $^{***}p<0.01$.
\end{tablenotes}
\end{table}

%% file: tables/robustness/p1_pre_post_regression.tex
\begin{table}[htbp]
\centering
\caption{Pre-Post OIS Volatility Relationship}
\label{tab:p1_pre_post_reg}
\begin{tabular}{lccc}
\toprule
 & \multicolumn{3}{c}{Post-Meeting OIS Volatility} \\
\cmidrule(lr){2-4}
 & (1) & (2) & (3) \\
\midrule
Pre-Meeting OIS Volatility & 0.433** & 0.323** & 0.291** \\
 & (0.183) & (0.153) & (0.147) \\[0.5em]
Synthetic SD & & 0.320*** & 0.314*** \\
 & & (0.056) & (0.059) \\[0.5em]
Constant & 0.042*** & 0.018*** & 0.016*** \\
 & (0.012) & (0.006) & (0.005) \\
\midrule
Maturity FE & No & No & Yes \\
Observations & 540 & 540 & 540 \\
$R^2$ & 0.248 & 0.379 & 0.403 \\
Adjusted $R^2$ & 0.247 & 0.376 & 0.398 \\
\bottomrule
\end{tabular}
\begin{tablenotes}
\scriptsize
\item \textbf{Note}: OLS regression of post-meeting OIS volatility (OIS min-max range 1 day after ECB press conference) on 1) pre-meeting volatility (average OIS min-max range in the 3 days before the conference) and 2) LLM-generated synthetic disagreement. Synthetic SD denotes the cross-sectional standard deviation of rate forecasts across 30 heterogeneous LLM-based agents interpreting ECB press conference transcripts, generated using Gemini 2.5-Flash with temperature = 1. Observations are pooled across three tenors (3-month, 2-year, and 10-year OIS) in Column (1) and (2). Heteroskedasticity-robust standard errors in parentheses. *** p$<$0.01, ** p$<$0.05, * p$<$0.10. 
\end{tablenotes}
\end{table}

%% file: tables/robustness/prompt_variations_table.tex
\begin{table}[H]
\centering
\begingroup\footnotesize
\caption{List of Prompt Variations} 
\label{tab:prompt_variations}
\begin{tabular}{llp{7cm}p{3cm}}
  \hline
Variation & Type & Key Changes & Modified Section \\ 
  \hline
  1 & Minor & Individual traders → distinct market participants & Context \\ 
    2 & Minor & Trading decision → investment choice & Context \\ 
    3 & Minor & Risk aversion order: High/Medium/Low → Low/Medium/High & Trader Characteristics \\ 
    4 & Minor & Interpretation of conference → analysis of meeting & Context \\ 
    5 & Minor & Markdown table → structured table format & Guidelines \\ 
    6 & Minor & Confidence score → certainty level & Task \\ 
    7 & Minor & Monetary policy decisions → interest rate policies & Context \\ 
    8 & Minor & Maximize profit → optimize returns & Context \\ 
    9 & Minor & Reordered behavioral biases (Anchoring first) & Trader Characteristics \\ 
   10 & Minor & Press conference → policy meeting throughout & Multiple sections \\ 
   \hline
 11 & Medium & Formal academic language; market agents & Context \\ 
   12 & Medium & Behavioral lenses; psychological profiles emphasis & Context \\ 
   13 & Medium & Risk tolerance: Conservative/Moderate/Aggressive & Trader Characteristics \& Context \\ 
   14 & Medium & Policy signals; forward guidance; processing approach & Context \& Characteristics \\ 
   15 & Medium & Position-taking behavior; different output headers & Task \& Output \\ 
   \hline
\end{tabular}
\endgroup
\end{table}

%% file: tables/llm_as_judge/judge_llm_prompts.tex
\begingroup
\scriptsize  
\setlength{\tabcolsep}{4pt}  

\begin{longtable}{@{}l r r p{11cm}@{}}  
\caption{Correlation by prompt iteration\label{tab:prompt_judge}}\\
\toprule
Iteration & Test Corr. & Train Corr. & Prompt \\
\midrule
\endfirsthead
\caption[]{Correlation by prompt iteration (continued)}\\
\toprule
Iteration & Test Corr. & Train Corr. & Prompt \\
\midrule
\endhead
\bottomrule
\endfoot

1 & 0.47 & 0.47 & 
Context: You are simulating the Euro area interest rate swap market with 30 traders interpreting ECB press conferences.  
Trader Characteristics: Risk Aversion (High/Medium/Low), Behavioral Biases (1–2 per trader), Interpretation Style (Fundamentalist, Sentiment Reader, Quantitative, Skeptic, Narrative-Driven).  
Task: Simulate each trader's trading action across 3 tenors (3M, 2Y, 10Y), with expected direction, new rate, and confidence.  
Output: Markdown table only: Date / Trader ID / Tenor / Expected Direction / New Expected Rate (\%). \\[2pt]

2 & 0.57 & 0.58 &
Context: Simulate Euro area swap market with 30 traders. Core Principle: Prediction dispersion must reflect transcript ambiguity (high ambiguity → wide spread, low → tight).  
Trader Characteristics same as above.  
Task: Output table per trader and tenor with expected direction and new expected rate. Markdown table only. \\[2pt]

3 & 0.37 & 0.31 &
Context: Primary goal: model market uncertainty. Two-step process: 1) Assess transcript ambiguity, 2) Calibrate dispersion (low:1–3bps, moderate:4–7bps, high:8–15+bps).  
Simulation: Generate 30 predictions per tenor per participant. Markdown table only. \\[2pt]

4 & 0.48 & 0.44 &
Context: Model uncertainty via natural dispersion from text interpretation. Process: 1) Deep textual analysis of clarity, ambiguity, surprise, 2) Generate 30 plausible interpretations (high ambiguity → high dispersion). Do not target specific std. Markdown table only. \\[2pt]

5 & 0.35 & 0.36 &
Context: Model distribution of market reactions. Goal: ensure prediction std reflects Clarity, Surprise, Conviction. Multi-factor assessment: Clarity of guidance, Surprise vs expectations, Conviction/hedging. Simulation: Generate 30 draws per tenor; draws may repeat in low-uncertainty cases. Markdown table only. \\

\end{longtable}
\endgroup